\documentclass[conference,10pt]{IEEEtran}

\usepackage{cite}

\ifCLASSINFOpdf
\else
   \usepackage[dvipdfmx]{graphicx}
\fi
\usepackage[cmex10]{amsmath}
\usepackage{amssymb}
\usepackage{enumitem}
\usepackage{balance}

\newtheorem{assumption}{Assumption}
\newtheorem{definition}{Definition}
\newtheorem{proposition}{Proposition}
\newtheorem{remark}{Remark}
\newtheorem{theorem}{Theorem}
\newtheorem{lemma}{Lemma}
\newcommand{\aeq}{\overset{\mathrm{a.s.}}{=}}
\newcommand{\ato}{\overset{\mathrm{a.s.}}{\to}}

\newcommand{\ageq}{\overset{\mathrm{a.s.}}{\geq}}

\begin{document}
%
\title{A Unified Framework of State Evolution\\for Message-Passing Algorithms}
%
%
%

\author{\IEEEauthorblockN{Keigo Takeuchi}
\IEEEauthorblockA{Dept.\ Electrical and Electronic Information Eng., 
Toyohashi University of Technology, 
Aichi 441-8580, Japan}
\IEEEauthorblockA{Email: takeuchi@ee.tut.ac.jp} 
}

\maketitle

\begin{abstract}
This paper presents a unified framework to understand the dynamics of 
message-passing algorithms in compressed sensing. State evolution is 
rigorously analyzed for a general error model that contains the error model 
of approximate message-passing (AMP), as well as that of orthogonal AMP. 
As a by-product, AMP is proved to converge asymptotically if the sensing matrix 
is orthogonally invariant and if the moment sequence of its asymptotic 
singular-value distribution coincide with that of the Mar\u{c}henko-Pastur 
distribution up to the order that is at most twice as large as the maximum 
number of iterations. 
\end{abstract}


%

\section{Introduction} 
Consider the recovery of an unknown $N$-dimensional signal vector 
$\boldsymbol{x}\in\mathbb{R}^{N}$ from an $M$-dimensional linear measurement 
vector $\boldsymbol{y}\in\mathbb{R}^{M}$, given by  
\begin{equation} \label{model}
\boldsymbol{y} 
= \boldsymbol{A}\boldsymbol{x} + \boldsymbol{w}. 
\end{equation}
In (\ref{model}), the sensing matrix $\boldsymbol{A}\in\mathbb{R}^{M\times N}$ 
is known, while the noise vector $\boldsymbol{w}\in\mathbb{R}^{M}$ is unknown. 
The purpose of this paper is to present a unified framework for analyzing the 
asymptotic performance of signal recovery via message-passing (MP). 

An important example of MP is approximate message-passing 
(AMP)~\cite{Donoho09}. Bayes-optimal AMP can be 
regarded as an {\em exact} approximation of belief 
propagation~\cite{Kabashima03} in the large-system limit---both $M$ and $N$ 
tend to infinity while the compression rate $\delta=M/N$ is kept ${\cal O}(1)$. 
Bayati {\em et al}.~\cite{Bayati11,Bayati15} analyzed the rigorous dynamics 
of AMP in the large system limit via state evolution (SE) when the sensing 
matrix $\boldsymbol{A}$ has independent and identically distributed (i.i.d.),  
zero-mean, and sub-Gaussian elements. Their result implies that, 
in spite of its low complexity, AMP can achieve the Bayes-optimal performance 
in a range of the compression rate $\delta$. 
However, AMP fails to converge when 
the sensing matrix is non-zero mean~\cite{Caltagirone14} or  
ill-conditioned~\cite{Rangan14}.    

Another important example of MP is orthogonal AMP (OAMP)~\cite{Ma17}. OAMP 
is also called vector AMP (VAMP)~\cite{Rangan17} and was originally 
proposed by Opper and Winther~\cite[Appendix D]{Opper05}. Bayes-optimal OAMP 
can be regarded as an large-system approximation of expectation 
propagation (EP)~\cite{Cespedes14,Takeuchi171}. The rigorous SE of OAMP 
was presented in the same conference when the sensing matrix is 
orthogonally invariant on the real field~\cite{Rangan17} or 
unitarily invariant on the complex field~\cite{Takeuchi171}. These rigorous 
results imply that OAMP converges for a wider class of sensing matrices than 
AMP because the class of orthogonally invariant matrices contains matrices 
with dependent elements. One disadvantage of OAMP is high complexity due to 
the requirement of one matrix inversion\footnote{
The singular-value decomposition (SVD) of $\boldsymbol{A}$ allows us to 
circumvent this requirement~\cite{Rangan17}. However, the SVD itself is high 
complexity, unless the sensing matrix has some special structure.} 
per iteration. See \cite{Takeuchi172} for a complexity reduction of OAMP. 

This paper proposes an SE framework for understanding both AMP and OAMP from 
a unified point of view. The proposed framework is based on a general recursive 
model of errors that contains the error models of both AMP and OAMP. The main 
point of the model is that the current errors depend on the whole history 
of errors in the preceding iterations, while the current errors in OAMP 
are determined only by the errors in the latest iteration. Under the assumption 
of orthogonally invariant sensing matrices, we present a rigorous SE 
analysis of the general error model in the large-system limit.  

The main contributions of this paper are twofold: One is the rigorous SE 
of the general error model that contains those of both AMP and OAMP. 
The result provides a framework for designing new MP algorithms 
that have the advantages of both AMP and OAMP~\cite{Cakmak17}: 
low complexity and the convergence property for orthogonally invariant 
sensing matrices. 

The other contribution is a detailed convergence analysis of AMP. AMP with 
the maximum number $T$ of iterations is proved to converge for orthogonally 
invariant sensing matrices if the moment sequence of the asymptotic  
eigenvalue (EV) distribution of $\boldsymbol{A}^{\mathrm{T}}\boldsymbol{A}$ 
coincides with that of 
the Mar\u{c}henko-Pastur distribution~\cite{Tulino04} up to order $2T$ at most. 
When $\boldsymbol{A}$ has i.i.d.\ zero-mean elements, the asymptotic EV 
distribution coincides with the Mar\u{c}henko-Pastur 
distribution perfectly. Thus, the i.i.d.\ assumption of $\boldsymbol{A}$ is 
too strong in guaranteeing the convergence of AMP, as long as a finite number 
of iterations are assumed. 

\section{Preliminaries}
\subsection{General Error Model}
Consider the singular-value decomposition (SVD) $\boldsymbol{A}
=\boldsymbol{U}\boldsymbol{\Sigma}\boldsymbol{V}^{\mathrm{T}}$ of the 
sensing matrix, in which 
$\boldsymbol{U}$ and $\boldsymbol{V}$ are $M\times M$ and $N\times N$ 
orthogonal matrices, respectively. 
We consider the following general error model in iteration~$t$: 
\begin{equation} \label{b}
\boldsymbol{b}_{t}
= \boldsymbol{V}^{\mathrm{T}}\tilde{\boldsymbol{q}}_{t}, 
\quad \tilde{\boldsymbol{q}}_{t} 
= \boldsymbol{q}_{t} - \sum_{t'=0}^{t-1}\langle 
 \partial_{t'}\boldsymbol{\psi}_{t-1}
\rangle \boldsymbol{h}_{t'}, 
\end{equation}
\begin{equation} \label{m}
\boldsymbol{m}_{t} 
= \boldsymbol{\phi}_{t}(\boldsymbol{b}_{0},\ldots,\boldsymbol{b}_{t}, 
\tilde{\boldsymbol{w}}), 
\end{equation}
\begin{equation} \label{h}
\boldsymbol{h}_{t} 
= \boldsymbol{V}\tilde{\boldsymbol{m}}_{t}, 
\quad \tilde{\boldsymbol{m}}_{t}
= \boldsymbol{m}_{t} - \sum_{t'=0}^{t}\langle
 \partial_{t'}\boldsymbol{\phi}_{t}
\rangle\boldsymbol{b}_{t'}, 
\end{equation}
\begin{equation} \label{q}
\boldsymbol{q}_{t+1} 
= \boldsymbol{\psi}_{t}(\boldsymbol{h}_{0},\ldots,\boldsymbol{h}_{t},
\boldsymbol{x}), 
\end{equation}
with $\tilde{\boldsymbol{w}}=\boldsymbol{U}^{\mathrm{T}}\boldsymbol{w}$ and 
the initial conditions 
$\boldsymbol{q}_{0}=\tilde{\boldsymbol{q}}_{0}=-\boldsymbol{x}$. 

In the general error model, the notation $\langle \boldsymbol{v}\rangle$ 
denotes the arithmetic mean $\langle \boldsymbol{v}\rangle=N^{-1}\sum_{n=1}^{N}
[\boldsymbol{v}]_{n}$ for $\boldsymbol{v}\in\mathbb{R}^{N}$. 
The functions 
$\boldsymbol{\phi}_{t}:
\mathbb{R}^{N\times(t+1)}\times\mathbb{R}^{M}\to\mathbb{R}^{N}$ and 
$\boldsymbol{\psi}_{t}:
\mathbb{R}^{N\times(t+2)}\to\mathbb{R}^{N}$ 
are the element-wise mapping of input vectors, i.e.\ 
\begin{equation} \label{phi}
[\boldsymbol{\phi}_{t}(\boldsymbol{b}_{0},\ldots,\boldsymbol{b}_{t},
\tilde{\boldsymbol{w}})]_{n} 
= \phi_{t,n}([\boldsymbol{b}_{0}]_{n},\ldots,[\boldsymbol{b}_{t}]_{n}, 
[\tilde{\boldsymbol{w}}]_{n}), 
\end{equation}
\begin{equation} \label{psi}
[\boldsymbol{\psi}_{t}(\boldsymbol{h}_{0},\ldots,\boldsymbol{h}_{t}, 
\boldsymbol{x})]_{n} 
= \psi_{t,n}([\boldsymbol{h}_{0}]_{n},\ldots,[\boldsymbol{h}_{t}]_{n}, 
[\boldsymbol{x}]_{n}) 
\end{equation}
for some functions $\phi_{t,n}:\mathbb{R}^{t+2}\to\mathbb{R}$ 
and $\psi_{t,n}:\mathbb{R}^{t+2}\to\mathbb{R}$. 
Finally, the notations $\partial_{t'}\boldsymbol{\phi}_{t}$ and 
$\partial_{t'}\boldsymbol{\psi}_{t}$ represent $N$-dimensional vectors of which 
the $n$th elements  $[\partial_{t'}\boldsymbol{\phi}_{t}]_{n}$ and 
$[\partial_{t'}\boldsymbol{\phi}_{t}]_{n}$ are given by  
the partial derivatives of $\phi_{t,n}$ and $\psi_{t,n}$ with respect to 
the $t'$th variable, respectively. 

The functions $\boldsymbol{\phi}_{t}$ and $\boldsymbol{\psi}_{t}$ may depend on 
the singular-values of the sensing matrix. Since the support of the 
asymptotic singular-value distribution of $\boldsymbol{A}$ is assumed 
to be compact in this paper, we do not write the dependencies of 
$\boldsymbol{\Sigma}$ explicitly. 

The general error model is composed of two systems with respect to 
$(\boldsymbol{b}_{t}, \boldsymbol{m}_{t})$ and $(\boldsymbol{h}_{t}, 
\boldsymbol{q}_{t+1})$, respectively. We refer to the former and latter 
systems as modules A and B, respectively.  

\begin{remark}
Suppose that the functions $\boldsymbol{\phi}_{t}$ and $\boldsymbol{\psi}_{t}$ 
depend only on the latest variables, i.e.\ 
$\boldsymbol{m}_{t}=\boldsymbol{\phi}_{t}
(\boldsymbol{b}_{t},\tilde{\boldsymbol{w}})$  and 
$\boldsymbol{q}_{t+1}=\boldsymbol{\psi}_{t}(\boldsymbol{h}_{t},\boldsymbol{x})$. 
Then, the general error model reduces to that of OAMP~\cite{Takeuchi171}. 
The functions $\boldsymbol{\phi}_{t}$ and $\boldsymbol{\psi}_{t}$ characterize 
the types of the linear filter and the thresholding function used in OAMP. 
Furthermore, the normalized squared norm $N^{-1}\|\boldsymbol{q}_{t+1}\|^{2}$ 
corresponds to the mean-square error (MSE) for the OAMP estimation of 
$\boldsymbol{x}$ in iteration~$t$. 
\end{remark}

\subsection{AMP}
We formulate an AMP error model similar to the general 
error model. 
Let $\boldsymbol{x}_{t}$ denote the AMP estimator of $\boldsymbol{x}$ in 
iteration~$t$. The update rules of AMP~\cite{Donoho09} are given by 
\begin{equation} \label{AMP_x}
\boldsymbol{x}_{t+1} 
= \boldsymbol{\theta}_{t}
(\boldsymbol{x}_{t} + \boldsymbol{A}^{\mathrm{T}}\boldsymbol{z}_{t}), 
\end{equation}
\begin{equation} \label{z}
\boldsymbol{z}_{t} 
= \boldsymbol{y} - \boldsymbol{A}\boldsymbol{x}_{t} 
+ \frac{\xi_{t-1}}{\delta}\boldsymbol{z}_{t-1},
\quad  
\xi_{t} 
= \left\langle
 \boldsymbol{\theta}_{t}'
 (\boldsymbol{x}_{t}+\boldsymbol{A}^{\mathrm{T}}\boldsymbol{z}_{t})
\right\rangle, 
\end{equation}
with $\boldsymbol{z}_{-1}=\boldsymbol{0}$ and 
$\boldsymbol{x}_{0}=\boldsymbol{0}$. In (\ref{AMP_x}), 
the thresholding function satisfies the separation condition 
$[\boldsymbol{\theta}_{t}(\boldsymbol{v})]_{n}
=\theta_{t}([\boldsymbol{v}]_{n})$ for $\boldsymbol{v}\in\mathbb{R}^{N}$ 
with a common scalar function $\theta_{t}:\mathbb{R}\to\mathbb{R}$. 

Let $\boldsymbol{h}_{t}=\boldsymbol{x}_{t}+\boldsymbol{A}^{\mathrm{T}}
\boldsymbol{z}_{t}-\boldsymbol{x}$ and $\boldsymbol{q}_{t+1}
=\boldsymbol{x}_{t+1}-\boldsymbol{x}$ denote the estimation errors before 
and after thresholding, respectively. From the definition~(\ref{q}), 
we find 
\begin{equation}
\boldsymbol{\psi}_{t}(\boldsymbol{h}_{t},\boldsymbol{x})
=\boldsymbol{\theta}_{t}(\boldsymbol{x}+\boldsymbol{h}_{t}) 
- \boldsymbol{x}.
\end{equation}
Then, the extrinsic vector $\tilde{\boldsymbol{q}}_{t}$ in (\ref{b}) 
for $t>0$ is given by 
\begin{equation} \label{q_tilde}
\tilde{\boldsymbol{q}}_{t}
= \boldsymbol{q}_{t} - \langle
 \boldsymbol{\theta}_{t-1}'(\boldsymbol{x}+\boldsymbol{h}_{t-1})
\rangle\boldsymbol{h}_{t-1} 
= \boldsymbol{q}_{t} - \xi_{t-1}\boldsymbol{h}_{t-1}.  
\end{equation}

To define the function $\boldsymbol{\phi}_{t}$ in (\ref{m}), we let 
\begin{equation} \label{AMP_h}
\boldsymbol{m}_{t}=\boldsymbol{V}^{\mathrm{T}}\boldsymbol{h}_{t}.
\end{equation} 
Substituting the definition of $\boldsymbol{h}_{t}$ yields 
\begin{equation} \label{AMP_m_tmp}
\boldsymbol{m}_{t} 
=\boldsymbol{V}^{\mathrm{T}}\boldsymbol{q}_{t} 
+ \boldsymbol{\Sigma}^{\mathrm{T}}\boldsymbol{U}^{\mathrm{T}}
\boldsymbol{z}_{t}
= \boldsymbol{b}_{t}
+ \xi_{t-1}\boldsymbol{m}_{t-1}
+ \boldsymbol{\Sigma}^{\mathrm{T}}\boldsymbol{U}^{\mathrm{T}}
\boldsymbol{z}_{t}, 
\end{equation}
with $\boldsymbol{b}_{t}=\boldsymbol{V}^{\mathrm{T}}\tilde{\boldsymbol{q}}_{t}$ 
and $\boldsymbol{m}_{-1}=\boldsymbol{0}$,  
where the second equality follows from (\ref{q_tilde}) and 
(\ref{AMP_h}). 
Left-multiplying (\ref{z}) by $\boldsymbol{\Sigma}^{\mathrm{T}}
\boldsymbol{U}^{\mathrm{T}}$ and using (\ref{model}), we obtain 
\begin{equation} \label{z_tilde}
\boldsymbol{\Sigma}^{\mathrm{T}}
\boldsymbol{U}^{\mathrm{T}}\boldsymbol{z}_{t} 
= - \boldsymbol{\Lambda}\boldsymbol{V}^{\mathrm{T}}\boldsymbol{q}_{t} 
+ \boldsymbol{\Sigma}^{\mathrm{T}}\tilde{\boldsymbol{w}}
+ \frac{\xi_{t-1}}{\delta}\boldsymbol{\Sigma}^{\mathrm{T}}
\boldsymbol{U}^{\mathrm{T}}\boldsymbol{z}_{t-1},
\end{equation}
with $\boldsymbol{\Lambda}=\boldsymbol{\Sigma}^{\mathrm{T}}
\boldsymbol{\Sigma}$. Applying (\ref{q_tilde}), (\ref{AMP_h}) and 
(\ref{AMP_m_tmp}) to (\ref{z_tilde}), we arrive at  
\begin{IEEEeqnarray}{rl}
\boldsymbol{m}_{t} 
=& (\boldsymbol{I}_{N} - \boldsymbol{\Lambda})\boldsymbol{b}_{t} 
- \frac{\xi_{t-1}}{\delta}\boldsymbol{b}_{t-1} 
+ \boldsymbol{\Sigma}^{\mathrm{T}}\tilde{\boldsymbol{w}} 
\nonumber \\ 
+ \xi_{t-1}&\left\{
 \left(
  1 + \frac{1}{\delta}
 \right)\boldsymbol{I}_{N}  
 - \boldsymbol{\Lambda}
\right\}\boldsymbol{m}_{t-1}
- \frac{\xi_{t-1}\xi_{t-2}}{\delta}\boldsymbol{m}_{t-2}, 
\label{AMP_m}
\end{IEEEeqnarray}
with $\boldsymbol{b}_{t}=\boldsymbol{0}$ and 
$\boldsymbol{m}_{t}=\boldsymbol{0}$ for $t<0$. The right-hand side (RHS) 
of (\ref{AMP_m}) defines the function $\boldsymbol{\phi}_{t}$ recursively. 
Note that $\boldsymbol{m}_{t}$ depends on all vectors 
$\{\boldsymbol{b}_{0},\ldots, \boldsymbol{b}_{t}\}$.  

The only difference between the general and AMP error models is in 
(\ref{h}) and (\ref{AMP_h}). Instead of $\tilde{\boldsymbol{m}}_{t}$, 
the vector $\boldsymbol{m}_{t}$ is used to define $\boldsymbol{h}_{t}$ 
in the AMP. We will prove $\langle\partial_{t'}\boldsymbol{\phi}_{t}\rangle
\aeq0$ in the second main theorem. 

\subsection{Assumptions}
We follow \cite{Bayati11} to postulate Lipschitz-continuous functions as   
$\boldsymbol{\phi}_{t}$ and $\boldsymbol{\psi}_{t}$ in the general error model. 

\begin{assumption} \label{assumption_Lipschitz}
$\phi_{t,n}$ and $\psi_{t,n}$ are Lipschitz-continuous. 
Furthermore, $\boldsymbol{\phi}_{t}
(\boldsymbol{b}_{0},\ldots,\boldsymbol{b}_{t}, 
\tilde{\boldsymbol{w}})$ and $\boldsymbol{\psi}_{t}
(\boldsymbol{h}_{0},\ldots,\boldsymbol{h}_{t},\boldsymbol{x})$ are not a 
linear combination of the first $t+1$ vectors plus some function of 
the last vector.
\end{assumption}

The latter assumption implies that $\tilde{\boldsymbol{q}}_{t}$ and 
$\tilde{\boldsymbol{m}}_{t}$ in (\ref{b}) and (\ref{h}) depend on 
$(\boldsymbol{h}_{0},\ldots,\boldsymbol{h}_{t})$ and 
$(\boldsymbol{b}_{0},\ldots,\boldsymbol{b}_{t})$, respectively. 

We assume the following moment conditions on $\boldsymbol{x}$ and 
$\boldsymbol{w}$ to guarantee the existence of the second moments of the 
variables in the general error model. 

\begin{assumption} \label{assumption_signal}
The signal vector $\boldsymbol{x}$ has independent elements with bounded 
$(4+\epsilon)$th moments for some $\epsilon>0$. 
\end{assumption}

\begin{assumption} \label{assumption_noise}
The noise vector $\boldsymbol{w}$ has bounded $(4+\epsilon)$th moments 
for some $\epsilon>0$ and satisfies 
$M^{-1}\|\boldsymbol{w}\|^{2}\ato\sigma^{2}$ as $M\to\infty$. 
\end{assumption}

We follow \cite{Rangan17,Takeuchi171} to postulate orthogonally invariant 
sensing matrices. 

\begin{assumption} \label{assumption_A}
The sensing matrix $\boldsymbol{A}$ is orthogonally invariant. 
More precisely, the orthogonal matrices $\boldsymbol{U}$ and $\boldsymbol{V}$ 
in the SVD $\boldsymbol{A}=\boldsymbol{U}\boldsymbol{\Sigma}
\boldsymbol{V}^{\mathrm{T}}$ is independent of the other random variables and 
Haar-distributed~\cite{Tulino04}. 
The empirical EV distribution of 
$\boldsymbol{A}^{\mathrm{T}}\boldsymbol{A}$ converges almost surely (a.s.) to 
an asymptotic distribution with a compact support in the large-system limit. 
\end{assumption}




\subsection{Mar\u{c}henko-Pastur Distribution}
We review the Mar\u{c}henko-Pastur distribution. 
Assume that the sensing matrix $\boldsymbol{A}\in\mathbb{R}^{M\times N}$ has 
independent zero-mean Gaussian elements with variance $1/M$. The $k$th moment 
$M^{-1}\mathrm{Tr}\{(\boldsymbol{A}\boldsymbol{A}^{\mathrm{T}})^{k}\}$ of 
the empirical EV distribution of $\boldsymbol{A}\boldsymbol{A}^{\mathrm{T}}$ 
converges a.s.\ to that of the Mar\u{c}henko-Pastur distribution 
in the large-system limit. Instead of presenting the Mar\u{c}henko-Pastur 
distribution explicitly, we characterize it via the $\eta$-transform 
$\eta:[0,\infty)\to(0, 1]$, defined as 
\begin{equation}
\eta(x) 
= \lim_{M=\delta N\to\infty}\frac{1}{M}\mathrm{Tr}\left\{
 (\boldsymbol{I}_{M} + x\boldsymbol{A}\boldsymbol{A}^{\mathrm{T}})^{-1}
\right\}. 
\end{equation} 
As shown in \cite[Eq.~(2.120)]{Tulino04}, the $\eta$-transform of the 
Mar\u{c}henko-Pastur distribution is the positive solution to 
\begin{equation} \label{eta_transform_tmp}
\eta = 1 - \frac{1}{\delta} + \frac{1}{\delta(1+x\eta)}. 
\end{equation}

The $\eta$-transform defines the Mar\u{c}henko-Pastur distribution  
uniquely because the distribution is uniquely determined by the Stieltjes 
transform, which is given via analytic continuation of 
the $\eta$-transform~\cite{Tulino04}.  

We need the asymptotic EV distribution of $\boldsymbol{A}^{\mathrm{T}}
\boldsymbol{A}$, rather than $\boldsymbol{A}
\boldsymbol{A}^{\mathrm{T}}$. Define the $\eta$-transform of 
$\boldsymbol{A}^{\mathrm{T}}\boldsymbol{A}$ as 
\begin{equation} \label{eta_transform_def}
\tilde{\eta}(x) 
= \lim_{M=\delta N\to\infty}\frac{1}{N}\mathrm{Tr}\left\{
 (\boldsymbol{I}_{N} + x\boldsymbol{A}^{\mathrm{T}}\boldsymbol{A})^{-1}
\right\}. 
\end{equation} 
Since $\boldsymbol{A}\boldsymbol{A}^{\mathrm{T}}$ and 
$\boldsymbol{A}^{\mathrm{T}}\boldsymbol{A}$ have identical positive eigenvalues, 
we find the relationship 
\begin{equation}
\tilde{\eta}(x) 
= \delta\eta(x) + 1 - \delta. 
\end{equation}
Substituting this into (\ref{eta_transform_tmp}) yields 
\begin{equation} \label{eta_transform}
\tilde{\eta} 
= \frac{\delta}{\delta + x(\tilde{\eta}+\delta-1)}. 
\end{equation}

It is possible to calculate the moment sequence of the asymptotic EV 
distribution of $\boldsymbol{A}^{\mathrm{T}}\boldsymbol{A}$ via the 
$\eta$-transform $\tilde{\eta}$. Since the $\eta$-transform is uniformly 
bounded for all $x\in[0, \infty)$, we use the eigen-decomposition 
$\boldsymbol{A}^{\mathrm{T}}\boldsymbol{A}=\boldsymbol{V}
\boldsymbol{\Lambda}\boldsymbol{V}^{\mathrm{T}}$  
and the definition~(\ref{eta_transform_def}) to obtain  
\begin{equation} \label{power_series} 
\tilde{\eta}(x) 
= \sum_{k=0}^{\infty}
(-x)^{k}\mu_{k},
\end{equation}
\begin{equation} \label{moment} 
\mu_{k}
=\lim_{M=\delta N\to\infty}\frac{1}{N}\mathrm{Tr}(\boldsymbol{\Lambda}^{k}). 
\end{equation}
This implies that the $k$th moment $\mu_{k}$ of the asymptotic EV distribution 
of $\boldsymbol{A}^{\mathrm{T}}\boldsymbol{A}$ is given via the $k$th 
derivative of the $\eta$-transform at the origin. Direct calculation of 
the derivatives based on (\ref{eta_transform}) yields 
$\mu_{0}=1$, $\mu_{1}=1$, and $\mu_{2}=1+\delta^{-1}$. 

\section{Main results}
\subsection{State Evolution}
We analyze the dynamics of the general error model in the large-system limit. 
Let 
\begin{IEEEeqnarray}{rl}
\boldsymbol{B}_{t}=&(\boldsymbol{b}_{0},\ldots,\boldsymbol{b}_{t-1})
\in\mathbb{R}^{N\times t}, \\
\tilde{\boldsymbol{M}}_{t}=&(\tilde{\boldsymbol{m}}_{0},\ldots,
\tilde{\boldsymbol{m}}_{t-1})\in\mathbb{R}^{N\times t}, \\ 
\boldsymbol{H}_{t}=&(\boldsymbol{h}_{0},\ldots,\boldsymbol{h}_{t-1})
\in\mathbb{R}^{N\times t}, \\
\tilde{\boldsymbol{Q}}_{t}
=&(\tilde{\boldsymbol{q}}_{0},\ldots,\tilde{\boldsymbol{q}}_{t-1})
\in\mathbb{R}^{N\times t}.
\end{IEEEeqnarray} 
Define the set 
$\mathfrak{E}_{t,t'}=\{\boldsymbol{B}_{t'}, \tilde{\boldsymbol{M}}_{t'},
\boldsymbol{H}_{t}, \tilde{\boldsymbol{Q}}_{t+1}, \boldsymbol{x}, 
\boldsymbol{w}, \boldsymbol{U}, \boldsymbol{\Sigma}\}$. The set 
$\mathfrak{E}_{t,t}$ contains the whole history of the estimation errors 
just before evaluating (\ref{b}) in iteration~$t$, as well as all random 
variables with the only exception of $\boldsymbol{V}$, while 
$\mathfrak{E}_{t,t+1}$ includes the whole history just before evaluating 
(\ref{h}). We use the conditioning technique by 
Bolthausen~\cite{Bolthausen14} to obtain the following theorem:  

\begin{theorem} \label{theorem_SE}
Postulate Assumptions~\ref{assumption_Lipschitz}--\ref{assumption_A}. 
For all $\tau=0,1,\ldots$ and $\tau'=0,\ldots,\tau$, the following properties 
hold for module~A in the large-system limit. 
\begin{enumerate}[label=(A-\alph*)]
\item \label{property_a_A}
Let $\boldsymbol{\beta}_{t}=(\tilde{\boldsymbol{Q}}_{t}^{\mathrm{T}}
\tilde{\boldsymbol{Q}}_{t})^{-1}\tilde{\boldsymbol{Q}}_{t}^{\mathrm{T}}
\tilde{\boldsymbol{q}}_{t}$, and $\tilde{\boldsymbol{q}}_{t}^{\perp}
=\boldsymbol{P}_{\tilde{\boldsymbol{Q}}_{t}}^{\perp}\tilde{\boldsymbol{q}}_{t}$, 
with $\boldsymbol{P}_{\tilde{\boldsymbol{Q}}_{t}}^{\perp}
=\boldsymbol{I}_{N}-\tilde{\boldsymbol{Q}}_{t}
(\tilde{\boldsymbol{Q}}_{t}^{\mathrm{T}}
\tilde{\boldsymbol{Q}}_{t})^{-1}\tilde{\boldsymbol{Q}}_{t}^{\mathrm{T}}$.  
For $\tau>0$, the vector $\boldsymbol{b}_{\tau}$ conditioned on 
$\mathfrak{E}_{\tau,\tau}$ is statistically equivalent to 
\begin{equation} \label{property_a1}
\boldsymbol{b}_{\tau}|_{\mathfrak{E}_{\tau,\tau}}
\sim \boldsymbol{B}_{\tau}\boldsymbol{\beta}_{\tau} 
+ \boldsymbol{B}_{\tau}\boldsymbol{o}(1) 
+ \tilde{\boldsymbol{M}}_{\tau}\boldsymbol{o}(1) 
+ \boldsymbol{\Phi}_{(\tilde{\boldsymbol{M}}_{\tau}, \boldsymbol{B}_{\tau})}^{\perp}
\boldsymbol{\omega}_{t}.    
\end{equation}
In (\ref{property_a1}), 
the notation $\boldsymbol{o}(1)$ denotes a finite-dimensional vector of 
which all elements are $o(1)$. For a matrix $\boldsymbol{M}$, the notation 
$\boldsymbol{\Phi}_{\boldsymbol{M}}^{\perp}$ represents the matrix that is composed 
of all left-singular vectors of $\boldsymbol{M}$ associated with zero singular 
values. $\boldsymbol{\omega}_{t}$ is independent of the other random variables, 
orthogonally invariant, and has bounded $(4+\epsilon)$th moments for 
some $\epsilon>0$ satisfying 
$\|\boldsymbol{\omega}_{t}\|^{2}=\|\tilde{\boldsymbol{q}}_{t}^{\perp}\|^{2}$.    
\item \label{property_b_A}
\begin{equation} \label{property_b1}
\frac{1}{N}\boldsymbol{b}_{\tau'}^{\mathrm{T}}\boldsymbol{b}_{\tau}
- \frac{1}{N}\tilde{\boldsymbol{q}}_{\tau'}^{\mathrm{T}}
\tilde{\boldsymbol{q}}_{\tau} \ato 0. 
\end{equation}
\item \label{property_c_A} 
Suppose that  
$\tilde{\boldsymbol{\phi}}_{\tau}(\boldsymbol{B}_{\tau+1},\tilde{\boldsymbol{w}})
:\mathbb{R}^{N\times(\tau+1)}\times\mathbb{R}^{M}
\to\mathbb{R}^{N}$ satisfies the separation condition like (\ref{phi}), and 
that each function $[\tilde{\boldsymbol{\phi}}_{\tau}]_{n}$ is 
Lipschitz-continuous. Then, 
\begin{equation} \label{property_c1}
\frac{1}{N}\boldsymbol{b}_{\tau'}^{\mathrm{T}}\left(
 \tilde{\boldsymbol{\phi}}_{\tau}
 - \sum_{t'=0}^{\tau}\left\langle
  \partial_{t'}\tilde{\boldsymbol{\phi}}_{\tau}
 \right\rangle\boldsymbol{b}_{t'}
\right)
\ato0. 
\end{equation}
\item \label{property_d_A}
There is some $C>0$ such that the minimum eigenvalue 
of $N^{-1}(\tilde{\boldsymbol{M}}_{\tau+1}^{\mathrm{T}}
\tilde{\boldsymbol{M}}_{\tau+1})^{-1}$ is a.s.\ larger than $C$. 
\end{enumerate}

For module~B, on the other hand, the following properties hold 
in the large-system limit: 
\begin{enumerate}[label=(B-\alph*)]
\item \label{property_a_B}
Let $\boldsymbol{\alpha}_{t}=(\tilde{\boldsymbol{M}}_{t}^{\mathrm{T}}
\tilde{\boldsymbol{M}}_{t})^{-1}\tilde{\boldsymbol{M}}_{t}^{\mathrm{T}}
\tilde{\boldsymbol{m}}_{t}$ and $\tilde{\boldsymbol{m}}_{t}^{\perp}
=\boldsymbol{P}_{\tilde{\boldsymbol{M}}_{t}}^{\perp}\tilde{\boldsymbol{m}}_{t}$.  
Then, the vector $\boldsymbol{h}_{\tau}$ conditioned on 
$\mathfrak{E}_{\tau,\tau+1}$ is statistically equivalent to 
\begin{equation} \label{property_a2}
\boldsymbol{h}_{0}|_{\mathfrak{E}_{0,1}}
\sim o(1)\boldsymbol{q}_{0}+ \boldsymbol{\Phi}_{\boldsymbol{q}_{0}}^{\perp}
\tilde{\boldsymbol{\omega}}_{0} 
\end{equation}
for $\tau=0$, otherwise 
\begin{IEEEeqnarray}{rl}
\boldsymbol{h}_{\tau}|_{\mathfrak{E}_{\tau,\tau+1}}
\sim& \boldsymbol{H}_{\tau}\boldsymbol{\alpha}_{\tau} 
+ \boldsymbol{H}_{\tau}\boldsymbol{o}(1) 
+ \tilde{\boldsymbol{Q}}_{\tau+1}\boldsymbol{o}(1) \nonumber \\
&+ \boldsymbol{\Phi}_{(\tilde{\boldsymbol{Q}}_{\tau+1},\boldsymbol{H}_{\tau})}^{\perp} 
\tilde{\boldsymbol{\omega}}_{t}.
\end{IEEEeqnarray}
In (\ref{property_a2}), 
$\tilde{\boldsymbol{\omega}}_{t}$ is an independent and orthogonally invariant 
vector, and has bounded $(4+\epsilon)$th moments for some $\epsilon>0$  
satisfying $\|\tilde{\boldsymbol{\omega}}_{0}\|^{2}
=\|\tilde{\boldsymbol{m}}_{0}\|^{2}$ and 
$\|\tilde{\boldsymbol{\omega}}_{t}\|^{2}
=\|\tilde{\boldsymbol{m}}_{t}^{\perp}\|^{2}$ for $t>0$.    
\item \label{property_b_B}
\begin{equation} \label{property_b2}
\frac{1}{N}\boldsymbol{h}_{\tau'}^{\mathrm{T}}\boldsymbol{h}_{\tau}
- \frac{1}{N}\tilde{\boldsymbol{m}}_{\tau'}^{\mathrm{T}}
\tilde{\boldsymbol{m}}_{\tau}
\ato 0. 
\end{equation}
\item \label{property_c_B} 
Suppose that  
$\tilde{\boldsymbol{\psi}}_{\tau}(\boldsymbol{H}_{\tau+1},\boldsymbol{x}):
\mathbb{R}^{N\times(\tau+2)}\to\mathbb{R}^{N}$ satisfies the separation condition 
like (\ref{psi}), and that each function 
$[\tilde{\boldsymbol{\psi}}_{\tau}]_{n}$ is Lipschitz-continuous. Then, 
\begin{equation} \label{property_c2}
\frac{1}{N}\boldsymbol{h}_{\tau'}^{\mathrm{T}}\left(
 \tilde{\boldsymbol{\psi}}_{\tau}
 - \sum_{t'=0}^{\tau}\left\langle
  \partial_{t'}\tilde{\boldsymbol{\psi}}_{\tau} 
 \right\rangle\boldsymbol{h}_{t'}
\right)
\ato0. 
\end{equation}
\item \label{property_d_B}
There is some $C>0$ such that the minimum eigenvalue of 
$N^{-1}(\tilde{\boldsymbol{Q}}_{\tau+2}^{\mathrm{T}}
\tilde{\boldsymbol{Q}}_{\tau+2})^{-1}$ are a.s.\ larger than $C$. 
\end{enumerate}
\end{theorem}
\begin{IEEEproof}
See Appendix~\ref{appen_proof_theorem_SE}. 
\end{IEEEproof}

Theorem~\ref{theorem_SE} was proved in \cite{Takeuchi171,Rangan17} 
for the case of functions $\boldsymbol{\phi}_{t}$ and 
$\boldsymbol{\psi}_{t}$ that depend only on   
$\boldsymbol{b}_{t}$ and $\boldsymbol{h}_{t}$, respectively. 
Theorem~\ref{theorem_SE} is a generalization of \cite{Takeuchi171,Rangan17} 
to the case of the general functions~(\ref{phi}) and 
(\ref{psi}). 

Properties~\ref{property_c_A} and \ref{property_c_B} imply the orthogonality 
between $\boldsymbol{b}_{\tau}$ and $\tilde{\boldsymbol{m}}_{t}$ and between 
$\boldsymbol{h}_{\tau}$ and $\tilde{\boldsymbol{q}}_{t+1}$ in the general error 
model. Thus, we refer to MP algorithms as long-memory OAMP (LM-OAMP) 
if their error models are contained in the general error model. 

If $\boldsymbol{q}_{t+1}$ corresponds to the estimation error of an MP 
algorithm in iteration~$t$, we need to evaluate the 
MSE $N^{-1}\|\boldsymbol{q}_{t+1}\|^{2}$ in the large-system 
limit. While Theorem~\ref{theorem_SE} allows us to analyze the MSE, this 
paper does not discuss any more analysis in the general error model.   
The MSE should be considered for each concrete MP algorithm. 

Because of space limitation, we have focused on a performance measure, 
such as MSE, that requires the existence of the second moments of the 
variables in the general error model. As considered in \cite{Bayati11}, 
it is straightforward to extend Theorem~\ref{theorem_SE} to the case of 
general performance measures in terms of pseudo-Lipschitz functions.  

\subsection{AMP} 
We next prove that the general error model contains the AMP error model 
under an assumption on the asymptotic EV distribution of 
$\boldsymbol{A}^{\mathrm{T}}\boldsymbol{A}$. 

\begin{theorem} \label{theorem_AMP}
Consider the AMP error model, 
postulate Assumptions~\ref{assumption_Lipschitz}--\ref{assumption_A}, 
and suppose that the moment sequence of the asymptotic EV distribution of 
$\boldsymbol{A}^{\mathrm{T}}\boldsymbol{A}$ coincides with that of the 
Mar\u{c}henko-Pastur distribution up to order~$T$. Then, 
$\tilde{\boldsymbol{m}}_{t}\aeq\boldsymbol{m}_{t}
+\boldsymbol{B}_{t+1}\boldsymbol{o}(1)$ holds for 
all $t<T$ in the large-system limit. 
\end{theorem}
\begin{IEEEproof}
See Section~\ref{sec4}. 
\end{IEEEproof}

The only difference between the general and AMP error models 
is in (\ref{h}) and (\ref{AMP_h}). Thus, Theorem~\ref{theorem_AMP}  
implies that the general error model contains the AMP error model 
in the large-system limit. As long as the number of iterations is finite, 
it should be possible to construct orthogonally invariant 
sensing matrices satisfying two conditions: One is that the sensing 
matrices have dependent elements. The other condition is that the moment 
sequence of the asymptotic EV distribution of 
$\boldsymbol{A}^{\mathrm{T}}\boldsymbol{A}$ is equal to that of the 
Mar\u{c}henko-Pastur distribution up to the required order. 
Thus, we conclude that Theorems~\ref{theorem_SE} and \ref{theorem_AMP} are 
the first rigorous result on the asymptotic dynamics of the AMP for 
non-independent sensing matrices.

\begin{remark}
Instead of evaluating $N^{-1}\boldsymbol{m}_{t}^{\mathrm{T}}\boldsymbol{m}_{t}$ 
directly, we present a sufficient condition for guaranteeing that  
the MSE $N^{-1}\|\boldsymbol{q}_{t+1}\|^{2}$ coincides with that for the case of 
zero-mean i.i.d.\ Gaussian sensing matrices~\cite{Bayati11}. 
From (\ref{AMP_m}), 
$N^{-1}\|\boldsymbol{m}_{t}\|^{2}$ depends on the 
asymptotic moments $\{\mu_{k}\}$ up to order $2t+2$. Thus, the MSE 
$N^{-1}\|\boldsymbol{q}_{t+1}\|^{2}$ coincides with that in \cite{Bayati11} for 
all $t<T$ in the large-system limit if the moment sequence of the 
asymptotic EV distribution of $\boldsymbol{A}^{\mathrm{T}}\boldsymbol{A}$ 
is equal to that of the Mar\u{c}henko-Pastur distribution up to order~$2T$. 
A future work is to analyze what occurs between the orders 
$T$ and $2T$.  
\end{remark}

\section{Proof of Theorem~\ref{theorem_AMP}}
\label{sec4}
Let $g_{\tau',\tau}^{(k)}=\langle\boldsymbol{\Lambda}^{k}
\partial_{\tau'}\boldsymbol{\phi}_{\tau}\rangle$ with $\boldsymbol{\phi}_{\tau}$ 
defined as the RHS of (\ref{AMP_m}). The goal is to prove 
$g_{\tau',\tau}^{(0)}\ato0$ for all $0\leq \tau<T$ and $0\leq \tau'\leq \tau$ 
in the large-system limit. 

The proof is by induction with respect to $\tau$. For $\tau=\tau'$, 
we use (\ref{AMP_m}) to obtain 
\begin{equation} \label{gtt}
g_{\tau,\tau}^{(k)} 
\aeq \mu_{k}-\mu_{k+1} + o(1)
\end{equation}
in the large-system limit, where the $k$th moment $\mu_{k}$ is 
defined in (\ref{moment}). 
In particular, for $\tau=0$ and $\tau=1$ we use $\mu_{0}=\mu_{1}=1$ to 
find $g_{\tau,\tau}^{(0)}\ato0$ in the large-system limit. 

Let $\tau=1$. Since we have proved $g_{0,0}^{(0)}\ato0$, we can use 
Property~\ref{property_a_B} for $\tau=0$. Thus, $\xi_{0}$ converges a.s.\ to 
a constant independent of $\boldsymbol{b}_{0}$ in the large-system limit. 
Using (\ref{AMP_m}) yields  
\begin{IEEEeqnarray}{rl}
\frac{g_{\tau-1,\tau}^{(k)}}{\xi_{\tau-1}} 
\aeq&- \frac{\mu_{k}}{\delta}
+ \left(
 1 + \frac{1}{\delta}
\right)g_{\tau-1,\tau-1}^{(k)}
- g_{\tau-1,\tau-1}^{(k+1)} + o(1)
\nonumber \\
\aeq& - \frac{\mu_{k+1}}{\delta}
+ g_{\tau-1,\tau-1}^{(k)}
- g_{\tau-1,\tau-1}^{(k+1)} + o(1) \label{gt-1t}
\end{IEEEeqnarray}
for $\tau=1$, where the second equality follows from the identity 
$\mu_{k}
\aeq g_{\tau-1,\tau-1}^{(k)} + \mu_{k+1} + o(1)$ obtained 
from (\ref{gtt}). Thus, we find $g_{0,1}^{(0)}/\xi_{0}\ato0$ 
in the large-system limit. 

Assume that there is some $t<T$ such that $g_{\tau',\tau}^{(k)}\ato0$ holds for 
all $0\leq \tau<t$ and $0\leq \tau'\leq \tau$. 
We prove $g_{\tau',t}^{(0)}\ato0$ for all $\tau'\leq t$.  
The induction hypothesis allows us to use Property~\ref{property_a_B} 
for all $\tau< t$, 
so that, for all $\tau<t$, $\xi_{\tau}$ converges a.s.\ to a constant 
independent of $\{\boldsymbol{b}_{0},\ldots,\boldsymbol{b}_{\tau}\}$ in the 
large-system limit. This observation implies that (\ref{gt-1t}) holds 
for all $\tau\leq t$. Furthermore, we use (\ref{AMP_m}) to obtain 
\begin{equation} \label{gt't}
\frac{g_{\tau',\tau}^{(k)}}{\xi_{\tau-1}} 
\aeq \left(
 1 + \frac{1}{\delta}
\right)g_{\tau',\tau-1}^{(k)}
- g_{\tau',\tau-1}^{(k+1)}
- \frac{\xi_{\tau-2}}{\delta}g_{\tau',\tau-2}^{(k)} + o(1) 
\end{equation} 
for all $\tau\leq t$ and $\tau'<\tau-1$. 

We simplify the recursive system (\ref{gtt}), (\ref{gt-1t}), and 
(\ref{gt't}). Let $g_{\tau',\tau}^{(k)}=a_{\tau}\tilde{g}_{\tau',\tau}^{(k)}/a_{\tau'}$, 
with $a_{0}=1$ and $a_{\tau}=\xi_{\tau-1}a_{\tau-1}$ for all $1\leq\tau\leq t$. 
Applying these definitions to (\ref{gtt}), (\ref{gt-1t}), and 
(\ref{gt't}), we have 
\begin{equation} \label{g0}
\tilde{g}_{\tau,\tau}^{(k)} 
\aeq \mu_{k} - \mu_{k+1} + o(1), 
\end{equation}
\begin{equation} \label{g1}
\tilde{g}_{\tau-1,\tau}^{(k)} 
\aeq - \frac{\mu_{k+1}}{\delta}
+ \tilde{g}_{\tau-1,\tau-1}^{(k)}
- \tilde{g}_{\tau-1,\tau-1}^{(k+1)} + o(1),  
\end{equation}
\begin{equation} \label{gt}
\tilde{g}_{\tau',\tau}^{(k)} 
\aeq \left(
 1 + \frac{1}{\delta}
\right)\tilde{g}_{\tau',\tau-1}^{(k)}
- \tilde{g}_{\tau',\tau-1}^{(k+1)}
- \frac{\tilde{g}_{\tau',\tau-2}^{(k)}}{\delta}
+ o(1). 
\end{equation} 
The simplified system~(\ref{g0})--(\ref{gt}) implies 
that $\tilde{g}_{\tau',\tau}^{(k)}$ is stationary with respect to 
$\tau'$ and $\tau$. In other words, 
$\tilde{g}_{\tau',\tau}^{(k)}$ depends on $\tau$ and $\tau'$ only through the 
difference $\tau-\tau'$. 

Let $g_{\tau}^{(k)}=\tilde{g}_{0,\tau}^{(k)}$ for $\tau\leq t$, which satisfies the 
recursive system~(\ref{g0}), (\ref{g1}), and (\ref{gt}) with 
$\tilde{g}_{\tau',\tau}^{(k)}$ replaced by $g_{\tau-\tau'}^{(k)}$.  
It is sufficient to prove $g_{t}^{(0)}\ato0$ in the large-system limit.  
By definition, $g_{\tau}^{(k)}$ is independent of the higher-order moments 
$\mu_{j}$ for all $j>\tau+k+1$. As long as 
$t<T$ is assumed, the sequence $\{g_{0}^{(0)},\ldots,g_{t}^{(0)}\}$ is 
determined by the moments up to order~$T$. Without loss of generality, 
we can assume that the asymptotic EV distribution of 
$\boldsymbol{A}^{\mathrm{T}}\boldsymbol{A}$ coincides with the 
Mar\u{c}henko-Pastur distribution perfectly. 

To prove $g_{t}^{(0)}\ato0$, we define the generating 
function of $\{g_{\tau}^{(k)}\}$ as 
\begin{equation} \label{generating_function} 
G(x,y) 
= \sum_{\tau=0}^{\infty}G_{\tau}(x)y^{\tau}, 
\end{equation} 
with 
\begin{equation} \label{generating_function_t}
G_{\tau}(x)
= \sum_{k=0}^{\infty}g_{\tau}^{(k)}x^{k} - \frac{g_{\tau-1}^{(0)}}{x}, 
\quad g_{-1}^{(0)}=0, 
\end{equation}
where $g_{\tau}^{(k)}$ satisfies the recursive system~(\ref{g0}), 
(\ref{g1}), and (\ref{gt}) with $\tilde{g}_{\tau',\tau}^{(k)}$ replaced by 
$g_{\tau-\tau'}^{(k)}$. Note that we have extended the definition of 
$g_{\tau}^{(k)}$ with respect to $\tau$ from $\{0,\ldots,t\}$ to 
all non-negative integers. From the induction hypothesis 
$g_{t-1}^{(0)}\ato0$, it is sufficient to prove $G_{t}(0)\ato0$. 

We first derive an explicit formula of $G(x,y)$. 
From (\ref{g0}), (\ref{g1}), and (\ref{gt}), we utilize the power-series 
representation~(\ref{power_series}) to obtain  
\begin{equation} \label{G0}
G_{0}(x) 
\aeq \tilde{\eta}(-x) - \frac{\tilde{\eta}(-x) - 1}{x}+ o(1), 
\end{equation}
\begin{equation} \label{G1}
G_{1}(x) 
\aeq - \frac{\tilde{\eta}(-x) - 1}{\delta x}
+ \left(
 1 - \frac{1}{x}
\right)G_{0}(x)
+ o(1), 
\end{equation}
\begin{equation} \label{Gt}
G_{\tau}(x) 
\aeq \left(
 1 + \frac{1}{\delta} - \frac{1}{x}
\right)G_{\tau-1}(x) 
- \frac{G_{\tau-2}(x)}{\delta} + o(1) 
\end{equation}
for all $\tau>1$, where we have used $\mu_{0}=1$. From (\ref{Gt}), we have 
\begin{IEEEeqnarray}{rl}
G(x,y) 
&\aeq G_{0}(x) + yG_{1}(x)
- \frac{y^{2}}{\delta}G(x,y) 
\nonumber \\
+& \left(
 1 + \frac{1}{\delta} - \frac{1}{x}
\right)y\{G(x,y)-G_{0}(x)\} + o(1).  
\end{IEEEeqnarray}
Solving this equation with (\ref{G0}) and (\ref{G1}), we arrive at 
\begin{equation}
G(x,y) = \frac{P(x,y)}{Q(x,y)} + o(1), 
\end{equation}
with 
\begin{equation}
P(x,y) = (\delta x - \delta -xy)\tilde{\eta}(-x) + \delta, 
\end{equation}
\begin{equation}
Q(x,y) 
= \delta y + (y-\delta)(y-1)x. 
\end{equation}

We next prove that the numerator $P(x,y)$ is divisible by 
the denominator $Q(x,y)$ for $y\in(0, \min\{1, \delta\})$. 
It is sufficient to prove that $P(-x^{*},y)=0$ holds for the zero $-x^{*}$ 
of $Q(x,y)$, given by 
\begin{equation}
x^{*} = \frac{\delta y}{(y-\delta)(y-1)}>0. 
\end{equation}
Calculating $P(-x^{*},y)$ yields 
\begin{equation}
P(-x^{*},y) 
= \delta\left(
 1 - \frac{\tilde{\eta}(x^{*})}{1-y}
\right). 
\end{equation}
Since the $\eta$-transform $\tilde{\eta}$ satisfies (\ref{eta_transform}), 
we have 
\begin{equation}
\left\{
 \tilde{\eta}(x^{*}) - \frac{y-\delta}{y}
\right\}
\left\{
 \tilde{\eta}(x^{*}) - (1-y)
\right\} = 0. 
\end{equation}
The positivity of the $\eta$-transform implies that the correct solution 
is $\tilde{\eta}(x^{*})=1-y$. Thus, we arrive at $P(-x^{*},y)=0$. 

Finally, we prove $g_{t}^{(0)}\ato0$. For $y\neq0$, 
we use $\tilde{\eta}(0)=1$ to find $\lim_{x\to0}G(x,y)\ato0$. 
Since we have proved that $G(x,y)$ is a polynomial for 
all $y\in(0, \min\{1, \delta\})$, from (\ref{generating_function}) 
we can conclude $\lim_{x\to0}G_{\tau}(x)\ato0$ for all $\tau$. In particular, 
we use (\ref{generating_function_t}) and the induction hypothesis 
$g_{t-1}^{(0)}\ato0$ to arrive at $g_{t}^{(0)}\ato0$. Thus, 
Theorem~\ref{theorem_AMP} holds. 

\section*{Acknowledgment}
The author was in part supported by the Grant-in-Aid 
for Scientific Research~(B) (JSPS KAKENHI Grant Number 18H01441), Japan.




\bibliographystyle{IEEEtran}
\bibliography{IEEEabrv,kt-isit2019}

\begin{thebibliography}{10}
\providecommand{\url}[1]{#1}
\csname url@samestyle\endcsname
\providecommand{\newblock}{\relax}
\providecommand{\bibinfo}[2]{#2}
\providecommand{\BIBentrySTDinterwordspacing}{\spaceskip=0pt\relax}
\providecommand{\BIBentryALTinterwordstretchfactor}{4}
\providecommand{\BIBentryALTinterwordspacing}{\spaceskip=\fontdimen2\font plus
\BIBentryALTinterwordstretchfactor\fontdimen3\font minus
  \fontdimen4\font\relax}
\providecommand{\BIBforeignlanguage}[2]{{%
\expandafter\ifx\csname l@#1\endcsname\relax
\typeout{** WARNING: IEEEtran.bst: No hyphenation pattern has been}%
\typeout{** loaded for the language `#1'. Using the pattern for}%
\typeout{** the default language instead.}%
\else
\language=\csname l@#1\endcsname
\fi
#2}}
\providecommand{\BIBdecl}{\relax}
\BIBdecl

\bibitem{Donoho09}
D.~L. Donoho, A.~Maleki, and A.~Montanari, ``Message-passing algorithms for
  compressed sensing,'' \emph{Proc. Nat. Acad. Sci.}, vol. 106, no.~45, pp.
  18\,914--18\,919, Nov. 2009.

\bibitem{Kabashima03}
Y.~Kabashima, ``A {CDMA} multiuser detection algorithm on the basis of belief
  propagation,'' \emph{J. Phys. A: Math. Gen.}, vol.~36, no.~43, pp.
  11\,111--11\,121, Oct. 2003.

\bibitem{Bayati11}
M.~Bayati and A.~Montanari, ``The dynamics of message passing on dense graphs,
  with applications to compressed sensing,'' \emph{{IEEE} Trans. Inf. Theory},
  vol.~57, no.~2, pp. 764--785, Feb. 2011.

\bibitem{Bayati15}
M.~Bayati, M.~Lelarge, and A.~Montanari, ``Universality in polytope phase
  transitions and message passing algorithms,'' \emph{Ann. Appl. Probab.},
  vol.~25, no.~2, pp. 753--822, Apr. 2015.

\bibitem{Caltagirone14}
F.~Caltagirone, L.~Zdeborov\'a, and F.~Krzakala, ``On convergence of
  approximate message passing,'' in \emph{Proc. 2014 IEEE Int. Symp. Inf.
  Theory}, Honolulu, HI, USA, Jul. 2014, pp. 1812--1816.

\bibitem{Rangan14}
S.~Rangan, P.~Schniter, and A.~Fletcher, ``On the convergence of approximate
  message passing with arbitrary matrices,'' in \emph{Proc. 2014 IEEE Int.
  Symp. Inf. Theory}, Honolulu, HI, USA, Jul. 2014, pp. 236--240.

\bibitem{Ma17}
J.~Ma and L.~Ping, ``Orthogonal {AMP},'' \emph{IEEE Access}, vol.~5, pp.
  2020--2033, Jan. 2017.

\bibitem{Rangan17}
S.~Rangan, P.~Schniter, and A.~K. Fletcher, ``Vector approximate message
  passing,'' in \emph{Proc. 2017 IEEE Int. Symp. Inf. Theory}, Aachen, Germany,
  Jun. 2017, pp. 1588--1592.

\bibitem{Opper05}
M.~Opper and O.~Winther, ``Expectation consistent approximate inference,''
  \emph{J. Mach. Learn. Res.}, vol.~6, pp. 2177--2204, Dec. 2005.

\bibitem{Cespedes14}
J.~C\'espedes, P.~M. Olmos, M.~S\'anchez-Fern\'andez, and F.~Perez-Cruz,
  ``Expectation propagation detection for high-order high-dimensional {MIMO}
  systems,'' \emph{{IEEE} Trans. Commun.}, vol.~62, no.~8, pp. 2840--2849, Aug.
  2014.

\bibitem{Takeuchi171}
K.~Takeuchi, ``Rigorous dynamics of expectation-propagation-based signal
  recovery from unitarily invariant measurements,'' in \emph{Proc. 2017 IEEE
  Int. Symp. Inf. Theory}, Aachen, Germany, Jun. 2017, pp. 501--505.

\bibitem{Takeuchi172}
K.~Takeuchi and C.-K. Wen, ``Rigorous dynamics of expectation-propagation
  signal detection via the conjugate gradient method,'' in \emph{Proc. 18th
  IEEE Int. Workshop Sig. Process. Advances Wirel. Commun.}, Sapporo, Japan,
  Jul. 2017, pp. 88--92.

\bibitem{Cakmak17}
B.~\c{C}akmak, M.~Opper, O.~Winther, and B.~H. Fleury, ``Dynamical functional
  theory for compressed sensing,'' in \emph{Proc. 2017 IEEE Int. Symp. Inf.
  Theory}, Aachen, Germany, Jun. 2017, pp. 2143--2147.

\bibitem{Tulino04}
A.~M. Tulino and S.~Verd\'{u}, \emph{Random Matrix Theory and Wireless
  Communications}.\hskip 1em plus 0.5em minus 0.4em\relax Hanover, MA, USA: Now
  Publishers Inc., 2004.

\bibitem{Bolthausen14}
E.~Bolthausen, ``An iterative construction of solutions of the {TAP} equations
  for the {Sherrington}-{Kirkpatrick} model,'' \emph{Commun. Math. Phys.}, vol.
  325, no.~1, pp. 333--366, Jan. 2014.

\bibitem{Lyons88}
R.~Lyons, ``Strong laws of large numbers for weakly correlated random
  variables,'' \emph{Michigan Math. J.}, vol.~35, no.~3, pp. 353--359, 1988.

\end{thebibliography}

\appendices 
\section{Proof of Theorem~\ref{theorem_SE}}
\label{appen_proof_theorem_SE} 
\subsection{Properties of Pseudo-Lipschitz Functions} 
We present the definition and basic properties of pseudo-Lipschitz 
functions~\cite{Bayati11}.
\begin{definition}
A function $f:\mathbb{R}^{t}\to\mathbb{R}$ is called pseudo-Lipschitz of 
order~$k$ if there are some constants $L>0$ and $k\in\mathbb{N}$ such that, 
for all $\boldsymbol{x}\in\mathbb{R}^{t}$ and  
$\boldsymbol{y}\in\mathbb{R}^{t}$,  
\begin{equation} \label{Lipschitz}
|f(\boldsymbol{x})-f(\boldsymbol{y})|
\leq L(1 + \|\boldsymbol{x}\|^{k-1} + \|\boldsymbol{y}\|^{k-1})
\|\boldsymbol{x}-\boldsymbol{y}\|. 
\end{equation}
\end{definition}

In proving the following propositions, we use the equivalence between norms 
on $\mathbb{R}^{t}$ for finite $t\in\mathbb{R}$, i.e.\ 
$C_{1}\|\cdot\|_{q}\leq \|\cdot\|_{p}\leq C_{2}\|\cdot\|_{q}$ for some 
constants $C_{1}, C_{2}>0$. Note that 
$\|\cdot\|_{2}$ is abbreviated as $\|\cdot\|$. 
\begin{proposition} \label{proposition1}
Let $f:\mathbb{R}^{t}\to\mathbb{R}$ denote any pseudo-Lipschitz 
function of order~$k$. Then, there is some constant $C>0$ such that 
$|f(\boldsymbol{x})|\leq L(1+\|\boldsymbol{x}\|^{k})$ for all 
$\boldsymbol{x}\in\mathbb{R}^{t}$. 
\end{proposition}
\begin{IEEEproof}
Since $f$ is pseudo-Lipschitz of order~$k$, there is some constant $L'>0$ 
such that $|f(\boldsymbol{x})|\leq |f(\boldsymbol{0})| 
+ L'(1+\|\boldsymbol{x}\|^{k-1})\|\boldsymbol{x}\|$ holds for all 
$\boldsymbol{x}\in\mathbb{R}^{t}$. For $\|\boldsymbol{x}\|<1$, 
we have $|f(\boldsymbol{x})|\leq |f(\boldsymbol{0})| + 2L'$. Otherwise, 
$|f(\boldsymbol{x})|\leq |f(\boldsymbol{0})| + 2L'\|\boldsymbol{x}\|^{k}$. 
Thus, there is some constant $L>0$ such that 
$|f(\boldsymbol{x})|\leq L(1+\|\boldsymbol{x}\|^{k})$ holds. 
\end{IEEEproof}

Proposition~\ref{proposition1} implies that any pseudo-Lipschitz function  
$f(\boldsymbol{x})$ of order~$k$ is ${\cal O}(\|\boldsymbol{x}\|^{k})$ as 
$\|\boldsymbol{x}\|\to\infty$, while $f(\boldsymbol{x})={\cal O}
(\|\boldsymbol{x}\|)$ holds for any Lipschitz-continuous function $f$.  

\begin{proposition} \label{proposition2} 
Let $\boldsymbol{x}\in\mathbb{R}^{t}$ denote a random vector with bounded 
$k$th absolute moments for some $k\in\mathbb{N}$. Suppose that 
a function $f:\mathbb{R}^{t}\to\mathbb{R}$ is pseudo-Lipschitz of order~$k$ 
and almost everywhere (a.e.) differentiable. Then, we have 
$\mathbb{E}[|f(\boldsymbol{x})|]<\infty$ 
and $\mathbb{E}[|\partial_{t'}f(\boldsymbol{x})|]<\infty$. 
\end{proposition}
\begin{IEEEproof}
Using Proposition~\ref{proposition1}, we obtain 
\begin{equation}
\mathbb{E}[|f(\boldsymbol{x})|] 
\leq C\left(
 1 + \mathbb{E}\left[
  \|\boldsymbol{x}\|^{k}
 \right]
\right) < \infty,
\end{equation}
where the boundedness follows from that of the $k$th absolute moments of 
$\boldsymbol{x}$. 

The boundedness 
$\mathbb{E}[\partial_{t'}f(\boldsymbol{x})]<\infty$ is also obtained by 
repeating the same argument, since (\ref{Lipschitz}) implies 
\begin{IEEEeqnarray}{rl}
|\partial_{t'}f(\boldsymbol{x})| 
=& \lim_{\Delta x\to0}\left|
 \frac{f(\boldsymbol{x}+\Delta x\boldsymbol{e}_{t'}) - f(\boldsymbol{x})}
 {\Delta x}  
\right| \nonumber \\
\leq& L\left(
 1 + 2\|\boldsymbol{x}\|^{k-1}
\right), 
\end{IEEEeqnarray}
where $\boldsymbol{e}_{t'}$ denotes the $t'$th column of $\boldsymbol{I}_{t}$. 
Thus, Proposition~\ref{proposition2} holds. 
\end{IEEEproof}

\begin{proposition} \label{proposition3} 
Suppose that $f:\mathbb{R}\to\mathbb{R}$ and $g:\mathbb{R}\to\mathbb{R}$ 
are pseudo-Lipschitz of orders~$k_{1}$ and $k_{2}$, respectively. Then, 
$h(\boldsymbol{x})=f(x_{1})g(x_{2})$ is pseudo-Lipschitz of 
order~$(k_{1}+k_{2})$. 
\end{proposition}
\begin{IEEEproof}
From the pseudo-Lipschitz properties, 
there are some constants $L_{\mathrm{f}}, L_{\mathrm{g}}>0$ such that 
\begin{equation} \label{f_bound}
|f(x_{1})-f(y_{1})|
\leq L_{\mathrm{f}}(1+|x_{1}|^{k_{1}-1}+|y_{1}|^{k_{1}-1})|x_{1}-y_{1}|, 
\end{equation}
\begin{equation} \label{g_bound}
|g(x_{2})-g(y_{2})|
\leq L_{\mathrm{g}}(1+|x_{2}|^{k_{2}-1}+|y_{2}|^{k_{2}-1})|x_{2}-y_{2}|. 
\end{equation}
Without loss of generality, we assume $|x_{1}|\geq |y_{1}|$ and 
$|x_{2}|\geq |y_{2}|$. 
Using the triangle inequality yields    
\begin{IEEEeqnarray}{rl}
&|f(x_{1})g(x_{2}) - f(y_{1})g(y_{2})|  \nonumber \\
\leq& |f(x_{1})-f(y_{1})||g(x_{2})| + |f(y_{1})||g(x_{2})-g(y_{2})|. 
\end{IEEEeqnarray}
Applying the upper bounds $|f(y_{1})|\leq C_{\mathrm{f}}(1+|y_{1}|^{k_{1}})$ and 
$|g(x_{2})|\leq C_{\mathrm{g}}(1+|x_{2}|^{k_{2}})$ for some constants 
$C_{\mathrm{f}}, C_{\mathrm{g}}>0$ obtained from Proposition~\ref{proposition1}, 
as well as (\ref{f_bound}) and (\ref{g_bound}), we obtain  
\begin{IEEEeqnarray}{rl}
&|f(x_{1})g(x_{2}) - f(y_{1})g(y_{2})|  \nonumber \\
\leq& C(1+|x_{1}|^{k_{1}-1})(1+|x_{2}|^{k_{2}})|x_{1}-y_{1}| 
\nonumber \\ 
&+ C(1+|x_{1}|^{k_{1}})(1+|x_{2}|^{k_{2}-1})|x_{2}-y_{2}|
\end{IEEEeqnarray} 
for some constant $C>0$, where we have used $|x_{1}|\geq |y_{1}|$ and 
$|x_{2}|\geq |y_{2}|$. 
Since $|x_{1}|^{k}|x_{2}|^{k'}\leq(|x_{1}|+|x_{2}|)^{k+k'}$ holds for all 
$k\geq0$ and $k'\geq0$, we have  
\begin{equation}
|f(x_{1})g(x_{2}) - f(y_{1})g(y_{2})|
< C\{1+\|\boldsymbol{x}\|_{1}^{k_{1}+k_{2}-1}\}
\|\boldsymbol{x}-\boldsymbol{y}\|_{1}. 
\end{equation}
Proposition~\ref{proposition3} follows from 
the equivalence between the norms $\|\cdot\|_{1}$ and $\|\cdot\|$ on 
$\mathbb{R}^{2}$. 
\end{IEEEproof}

\subsection{Key Lemmas}
We present three key lemmas used in proving Theorem~\ref{theorem_SE}. 
\begin{lemma}[\cite{Rangan17}]  \label{lemma_conditioning} 
Suppose that the $N\times N$ orthogonal matrix $\boldsymbol{V}$ 
is Haar-distributed. 
For $0<t<N$, consider a noiseless linear measurement 
$\boldsymbol{Y}\in\mathbb{R}^{N\times t}$ of the unknown {\em signal} matrix 
$\boldsymbol{V}$ given by 
\begin{equation} \label{constraint}
\boldsymbol{Y}=\boldsymbol{V}\boldsymbol{X}, 
\end{equation}
where the known {\em sensing} matrix $\boldsymbol{X}\in\mathbb{R}^{N\times t}$ 
is full rank. Then, the posterior distribution of $\boldsymbol{V}$ 
given $\boldsymbol{X}$ and $\boldsymbol{Y}$ is statistically equivalent to 
\begin{equation}
\boldsymbol{V}|_{\boldsymbol{X}, \boldsymbol{Y}} 
\sim \boldsymbol{Y}(\boldsymbol{X}^{\mathrm{T}}\boldsymbol{X})^{-1}
\boldsymbol{X}^{\mathrm{T}} + \boldsymbol{\Phi}_{\boldsymbol{Y}}^{\perp}
\tilde{\boldsymbol{V}}(\boldsymbol{\Phi}_{\boldsymbol{X}}^{\perp})^{\mathrm{T}}, 
\end{equation}
where the $(N-t)\times(N-t)$ orthogonal matrix 
$\tilde{\boldsymbol{V}}$ is Haar-distributed. 
\end{lemma}

Lemma~\ref{lemma_conditioning} is the main lemma in the conditioning 
technique by Bolthausen~\cite{Bolthausen14}. The lemma is used to prove 
Properties~\ref{property_a_A} and \ref{property_a_B}. 

\begin{lemma} \label{lemma_Stein}
Let $\boldsymbol{z}=(z_{1},\ldots,z_{t})^{\mathrm{T}}\sim\mathcal{N}
(\boldsymbol{0},\boldsymbol{\Sigma})$. For all $k\in\mathbb{N}$, any 
pseudo-Lipschitz of order~$k$ and a.e.\  differentiable function 
$f:\mathbb{R}^{t}\to\mathbb{R}$ satisfies 
\begin{equation} \label{Stein_eq}
\mathbb{E}[z_{1}f(\boldsymbol{z})]
= \sum_{t'=1}^{t}\mathbb{E}[z_{1}z_{t'}]
\mathbb{E}\left[
 \partial_{t'}f(\boldsymbol{z})
\right].  
\end{equation}
\end{lemma}
\begin{IEEEproof} 
Proposition~\ref{proposition2} implies that both sides in (\ref{Stein_eq}) 
are bounded. 
For the eigen-decomposition $\boldsymbol{\Sigma}=\boldsymbol{U}
\boldsymbol{\Lambda}\boldsymbol{U}^{\mathrm{T}}$, we use the change of 
variables $\tilde{\boldsymbol{z}}=\boldsymbol{U}^{\mathrm{T}}\boldsymbol{z}$ 
to obtain 
\begin{equation}
\mathbb{E}[z_{1}f(\boldsymbol{z})]
= \sum_{\tau=1}^{t}U_{1\tau}\mathbb{E}[\tilde{z}_{\tau}f(\boldsymbol{U}
\tilde{\boldsymbol{z}})]. 
\end{equation}
Since $\tilde{\boldsymbol{z}}$ has independent elements, 
Stein's lemma implies 
\begin{IEEEeqnarray}{rl}
\mathbb{E}[z_{1}f(\boldsymbol{z})]
=& \sum_{\tau=1}^{t}U_{1\tau}\mathbb{E}[\tilde{z}_{\tau}^{2}]
\mathbb{E}\left[
 \frac{\partial f}{\partial \tilde{z}_{\tau}}
 (\boldsymbol{U}\tilde{\boldsymbol{z}})
\right] \nonumber \\
=& \sum_{\tau=1}^{t}U_{1\tau}[\boldsymbol{\Lambda}]_{\tau\tau}
\sum_{t'=1}^{t}U_{t' \tau}\mathbb{E}\left[
 \partial_{t'}f(\boldsymbol{z})
\right]. 
\end{IEEEeqnarray} 
Using the definition $[\boldsymbol{\Sigma}]_{1t'} 
=\sum_{\tau=1}^{t}U_{1\tau}[\boldsymbol{\Lambda}]_{\tau\tau}U_{t' \tau}$, 
we arrive at Lemma~\ref{lemma_Stein}. 
\end{IEEEproof}

Lemma~\ref{lemma_Stein} is used to prove Properties~\ref{property_c_A} and 
\ref{property_c_B}. The expression of the so-called Onsager terms---the 
second terms on $\tilde{\boldsymbol{q}}_{t}$ and $\tilde{\boldsymbol{m}}_{t}$ 
given in (\ref{b}) and (\ref{h})---originates from Lemma~\ref{lemma_Stein}.  

\begin{lemma} \label{lemma_SLLN}
Suppose that scalar functions $\{f_{n}:\mathbb{R}\to\mathbb{R}\}$ are 
pseudo-Lipschitz of order~$k$ for some $k\in\mathbb{N}$, that 
$\boldsymbol{a}\in\mathbb{R}^{N-t}$ is an 
orthogonally invariant vector with bounded $(2k+\epsilon)$th moments for 
some $\epsilon>0$, and that the limit 
$\lim_{N\to\infty}N^{-1}\|\boldsymbol{a}\|^{2}\aeq v>0$ holds 
for fixed $t\geq0$. Let $\boldsymbol{b}\in\mathbb{R}^{N-t}$ denote a 
deterministic vector satisfying $N^{-1}\|\boldsymbol{b}\|^{2}\to0$ and 
$N^{-1}\sum_{n=1}^{N}b_{n}^{2k-2}<\infty$. 
Let $\tilde{\boldsymbol{a}}=\boldsymbol{\Phi}^{\perp}\boldsymbol{a}$ for 
any $N\times(N-t)$ matrix $\boldsymbol{\Phi}^{\perp}$ with 
orthonormal columns. Then, we have 
\begin{equation}
\lim_{N\to\infty}\frac{1}{N}\sum_{n=1}^{N}\left\{
 f_{n}(\tilde{a}_{n}+b_{n})
 - \mathbb{E}\left[
  f_{n}(\sqrt{v}z_{n})
 \right]
\right\}
\aeq 0, 
\end{equation}
with $\boldsymbol{z}\sim\mathcal{N}(\boldsymbol{0}, \boldsymbol{I}_{N})$. 
\end{lemma}
\begin{IEEEproof} 
See Appendix~\ref{appen_proof_lemma_SLLN}. 
\end{IEEEproof}

Lemma~\ref{lemma_SLLN} is used to prove Properties \ref{property_c_A} and 
\ref{property_c_B}. The moment conditions of $\boldsymbol{x}$ and 
$\boldsymbol{w}$ in Assumptions~\ref{assumption_signal} and 
\ref{assumption_noise} are required for utilizing this lemma. 

\subsection{Properties in Module~A for $\tau=0$}
The proof of Theorem~\ref{theorem_SE} is by induction. We first prove 
the properties in Module~A for $\tau=0$. We need to prove 
Properties~\ref{property_b_A}, \ref{property_c_A}, and \ref{property_d_A} 
for $\tau=0$. 
We only prove Property~\ref{property_c_A} for $\tau=0$ 
since Properties~\ref{property_b_A} and \ref{property_d_A} are trivial 
for $\tau=0$. 
From the definition~(\ref{b}) and Assumption~\ref{assumption_signal}, 
$\boldsymbol{b}_{0}$ is orthogonally invariant and has 
bounded $(4+\epsilon)$th moments for some $\epsilon>0$. 
Furthermore, $\tilde{\boldsymbol{w}}=\boldsymbol{U}^{\mathrm{T}}\boldsymbol{w}$ 
is orthogonally invariant and has bounded $(4+\epsilon)$th moments 
from Assumption~\ref{assumption_noise}. Note that 
$b\tilde{\phi}_{0,n}(b, \tilde{w}_{n})$ is 
pseudo-Lipschitz of order $2$ from Proposition~\ref{proposition3}. 
Thus, we can use Lemma~\ref{lemma_SLLN}. 

Let $\boldsymbol{z}_{0}\sim\mathcal{N}(\boldsymbol{0},v_{0}\boldsymbol{I}_{N})$ 
with $v_{0}=\lim_{M=\delta N\to\infty}N^{-1}\|\boldsymbol{b}_{0}\|^{2}$. 
Using Lemma~\ref{lemma_SLLN} conditioned on $\tilde{\boldsymbol{w}}$ and 
then using the same lemma again, we obtain  
\begin{IEEEeqnarray}{rl}
\frac{1}{N}\boldsymbol{b}_{0}^{\mathrm{T}}
\tilde{\boldsymbol{\phi}}_{0}(\boldsymbol{b}_{0},\tilde{\boldsymbol{w}})
\aeq& \frac{1}{N}\mathbb{E}\left[
 \boldsymbol{z}_{0}^{\mathrm{T}}\tilde{\boldsymbol{\phi}}_{0}
 (\boldsymbol{z}_{0},\tilde{\boldsymbol{w}})
\right] + o(1) \nonumber \\
=& v_{0}\mathbb{E}\left[
 \left\langle
  \partial_{0}\tilde{\boldsymbol{\phi}}_{0}
  (\boldsymbol{z}_{0},\tilde{\boldsymbol{w}})
 \right\rangle
\right] + o(1) \nonumber \\
\aeq& \frac{1}{N}\boldsymbol{b}_{0}^{\mathrm{T}}
\boldsymbol{b}_{0}\left\langle
 \partial_{0}\tilde{\boldsymbol{\phi}}_{0}
 (\boldsymbol{b}_{0},\tilde{\boldsymbol{w}})
\right\rangle
 + o(1). \label{bm0}
\end{IEEEeqnarray}
where the second equality follows from Lemma~\ref{lemma_Stein}. 
For the last equality, we need a careful discussion: Lemma~\ref{lemma_SLLN} 
implies that the empirical distribution of 
$(\boldsymbol{b}_{0}, \tilde{\boldsymbol{w}})$ 
converges weakly to the distribution of $(\boldsymbol{z}_{0}, 
\tilde{\boldsymbol{w}})$ in the large-system limit. We use 
\cite[Lemma 5]{Bayati11} to obtain the last equality.  
Thus, Property~\ref{property_c_A} holds for $\tau=0$.

\subsection{Properties in Module~B for $\tau=0$}
Since Property~\ref{property_b_B} is trivial for $\tau=0$, we only prove 
the other properties in Module~B for $\tau=0$. 
We first prove Property~\ref{property_a_B} for $\tau=0$. Using 
Lemma~\ref{lemma_conditioning} with $\boldsymbol{Y}=\boldsymbol{q}_{t}$ 
and $\boldsymbol{X}=\boldsymbol{b}_{t}$ for (\ref{h}) yields 
\begin{equation}
\boldsymbol{h}_{0}|_{\mathfrak{E}_{0,1}}
\sim \frac{\boldsymbol{b}_{0}^{\mathrm{T}}
\tilde{\boldsymbol{m}}_{0}}{\|\boldsymbol{b}_{0}\|^{2}}
\boldsymbol{q}_{0} 
+ \boldsymbol{\Phi}_{\boldsymbol{q}_{0}}^{\perp}\tilde{\boldsymbol{\omega}}_{0} 
\aeq o(1)\boldsymbol{q}_{0} 
+ \boldsymbol{\Phi}_{\boldsymbol{q}_{0}}^{\perp}\tilde{\boldsymbol{\omega}}_{0},
\end{equation}
with $\tilde{\boldsymbol{\omega}}_{0}
=\tilde{\boldsymbol{V}}
(\boldsymbol{\Phi}_{\boldsymbol{b}_{0}}^{\perp})^{\mathrm{T}}
\tilde{\boldsymbol{m}}_{0}$, in which 
the last equality follows from (\ref{h}) and (\ref{property_c1}) for $\tau=0$. 
Note that $\tilde{\boldsymbol{\omega}}_{0}$ is an orthogonally invariant 
vector. Since $\boldsymbol{b}_{0}$ has bounded $(4+\epsilon)$th moments, 
from (\ref{m}), (\ref{h}), Assumption~\ref{assumption_Lipschitz}, and 
Assumption~\ref{assumption_noise}, $\tilde{\boldsymbol{m}}_{0}$ is so. 
Thus, $\tilde{\boldsymbol{\omega}}_{0}$ has bounded $(4+\epsilon)$th moments. 
Furthermore, we have $\|\tilde{\boldsymbol{\omega}}_{0}\|^{2}
=\tilde{\boldsymbol{m}}_{0}^{\mathrm{T}}\boldsymbol{P}_{\boldsymbol{b}_{0}}^{\perp}
\tilde{\boldsymbol{m}}_{0}=\|\tilde{\boldsymbol{m}}_{0}\|^{2}+o(N)$, because 
of (\ref{property_c1}) for $\tau=0$. Thus. Property~\ref{property_a_B} 
holds for $\tau=0$. 

We next prove Property~\ref{property_c_B} for $\tau=0$. 
Let $\tilde{\boldsymbol{z}}_{0}\sim\mathcal{N}(\boldsymbol{0}, 
\tilde{v}_{0}\boldsymbol{I}_{N})$ with 
$\tilde{v}_{0}=\lim_{M=\delta N\to\infty}N^{-1}\|\tilde{\boldsymbol{m}}_{0}\|^{2}$. 
Using Property~\ref{property_a_B} and Lemma~\ref{lemma_SLLN} yields 
\begin{IEEEeqnarray}{rl}
\frac{1}{N}\boldsymbol{h}_{0}^{\mathrm{T}}
\tilde{\boldsymbol{\psi}}_{0}(\boldsymbol{h}_{0},\boldsymbol{x}) 
\aeq& \frac{1}{N}\mathbb{E}\left[
 \tilde{\boldsymbol{z}}_{0}^{\mathrm{T}}\tilde{\boldsymbol{\psi}}_{0}
 (\tilde{\boldsymbol{z}}_{0}, \boldsymbol{x})
\right] + o(1) \nonumber \\
=& \tilde{v}_{0}\mathbb{E}\left[
 \left\langle 
  \partial_{0}\tilde{\boldsymbol{\psi}}_{0}(\tilde{\boldsymbol{z}}_{0}, 
  \boldsymbol{x})
 \right\rangle
\right] + o(1) \nonumber \\
\aeq& \frac{\|\boldsymbol{h}_{0}\|^{2}}{N}\left\langle
 \partial_{0}\tilde{\boldsymbol{\psi}}_{0}(\boldsymbol{h}_{0}, 
 \boldsymbol{x})
\right\rangle + o(1),  \label{hq0} 
\end{IEEEeqnarray}
where the second inequality is due to Lemma~\ref{lemma_Stein}, and 
where the last inequality follows from the definition of $\tilde{v}_{0}$, 
(\ref{h}), and the same argument as in the derivation of (\ref{bm0}). 
Thus, Property~\ref{property_c_B} holds for $\tau=0$.  

Finally, we prove Property~\ref{property_d_B} for $\tau=0$. 
From \cite[Lemmas 8 and 9]{Bayati11}, it is sufficient to prove 
$N^{-1}\|\tilde{\boldsymbol{q}}_{1}^{\perp}\|^{2}$ converges a.s.\ to a strictly 
positive constant in the large-system limit. By definition, we have 
\begin{IEEEeqnarray}{rl}
\frac{\|\tilde{\boldsymbol{q}}_{1}^{\perp}\|^{2}}{N} 
=& \frac{\|\tilde{\boldsymbol{q}}_{1}\|^{2}}{N} 
- \frac{(N^{-1}\tilde{\boldsymbol{q}}_{0}^{\mathrm{T}}
\tilde{\boldsymbol{q}}_{1})^{2}}
{N^{-1}\|\tilde{\boldsymbol{q}}_{0}\|^{2}} 
\nonumber \\
\aeq& \frac{\mathbb{E}[\|\tilde{\boldsymbol{q}}_{1}\|^{2}]}{N} 
- \frac{(N^{-1}\mathbb{E}[\tilde{\boldsymbol{q}}_{0}^{\mathrm{T}}
\tilde{\boldsymbol{q}}_{1}])^{2}}
{N^{-1}\mathbb{E}[\|\tilde{\boldsymbol{q}}_{0}\|^{2}]} 
+ o(1),  \label{q0_perp}
\end{IEEEeqnarray}
with 
\begin{equation}
\tilde{\boldsymbol{q}}_{1} 
= \boldsymbol{\psi}_{0}(\tilde{\boldsymbol{z}}_{0},\boldsymbol{x})
- \mathbb{E}\left[
 \langle\partial_{0}\boldsymbol{\psi}_{0}
 (\tilde{\boldsymbol{z}}_{0},\boldsymbol{x})\rangle
\right]\tilde{\boldsymbol{z}}_{0}, 
\end{equation}
where the second equality is obtained by repeating the same argument as 
in the derivation of (\ref{hq0}). 

In order to lower-bound (\ref{q0_perp}), we use the Cauchy-Schwarz 
inequality twice, 
\begin{IEEEeqnarray}{rl}
\left(
 \mathbb{E}[\tilde{\boldsymbol{q}}_{0}^{\mathrm{T}}
 \tilde{\boldsymbol{q}}_{1}]
\right)^{2}
=& \left(
 \mathbb{E}\left\{
  \tilde{\boldsymbol{q}}_{0}^{\mathrm{T}}
  \mathbb{E}_{\tilde{\boldsymbol{z}}_{0}}[\tilde{\boldsymbol{q}}_{1}]
 \right\}
\right)^{2} 
\leq \left(
 \mathbb{E}\left\{
  \|\tilde{\boldsymbol{q}}_{0}\|
  \|\mathbb{E}_{\tilde{\boldsymbol{z}}_{0}}[\tilde{\boldsymbol{q}}_{1}]\|
 \right\}
\right)^{2} \nonumber \\
\leq& \mathbb{E}[\|\tilde{\boldsymbol{q}}_{0}\|^{2}]
\mathbb{E}\{\|\mathbb{E}_{\tilde{\boldsymbol{z}}_{0}}
[\tilde{\boldsymbol{q}}_{1}]\|^{2}\}.  
\end{IEEEeqnarray} 
Substituting this upper bound into (\ref{q0_perp}) yields 
\begin{equation}
\frac{\|\tilde{\boldsymbol{q}}_{1}^{\perp}\|^{2}}{N} 
\ageq \frac{1}{N}\mathbb{E}\left\{
 \mathbb{E}_{\tilde{\boldsymbol{z}}_{0}}[\|\tilde{\boldsymbol{q}}_{1}\|^{2}] 
 - \|\mathbb{E}_{\tilde{\boldsymbol{z}}_{0}}
[\tilde{\boldsymbol{q}}_{1}]\|^{2}
\right\} + o(1), 
\end{equation}
which is strictly positive from Assumption~\ref{assumption_Lipschitz}. 
Thus, Property~\ref{property_d_B} holds for $\tau=0$. 

\subsection{Properties in Module A by Induction} \label{Module_A_induction}
Suppose that Theorem~\ref{theorem_SE} is correct for all $\tau<t$. 
We first prove Property~\ref{property_a_A} for $\tau=t$. 
The orthogonal matrix $\boldsymbol{V}^{\mathrm{T}}$ conditioned on 
$\mathfrak{E}_{t,t}$ satisfies the constraint 
\begin{equation}
(\tilde{\boldsymbol{M}}_{t}, \boldsymbol{B}_{t}) 
= \boldsymbol{V}^{\mathrm{T}}(\boldsymbol{H}_{t}, \tilde{\boldsymbol{Q}}_{t}). 
\end{equation}

We confirm that $(\boldsymbol{H}_{t}, \tilde{\boldsymbol{Q}}_{t})$ is full 
rank. The induction hypothesis~\ref{property_c_B} for all $\tau<t$ implies 
the orthogonality 
$N^{-1}\boldsymbol{h}_{\tau'}^{\mathrm{T}}\tilde{\boldsymbol{q}}_{\tau''}\aeq0$ 
for all $\tau''\leq t$ and $\tau'<\tau''$. Thus, we have 
\begin{equation}
(\boldsymbol{H}_{t}, \tilde{\boldsymbol{Q}}_{t})^{\mathrm{T}}
(\boldsymbol{H}_{t}, \tilde{\boldsymbol{Q}}_{t})
\aeq \begin{pmatrix}
\tilde{\boldsymbol{M}}_{t}^{\mathrm{T}}\tilde{\boldsymbol{M}}_{t} 
& \boldsymbol{O} \\
\boldsymbol{O} & \tilde{\boldsymbol{Q}}_{t}^{\mathrm{T}}\tilde{\boldsymbol{Q}}_{t}
\end{pmatrix} + o(N),
\end{equation}
where we have used the definition~(\ref{h}). The induction 
hypotheses~\ref{property_d_A} and \ref{property_d_B} imply that 
$(\boldsymbol{H}_{t}, \tilde{\boldsymbol{Q}}_{t})$ is full rank. 
Thus, we can use Lemma~\ref{lemma_conditioning} to obtain 
\begin{equation}
\boldsymbol{V}^{\mathrm{T}}\tilde{\boldsymbol{q}}_{t}|_{\mathfrak{E}_{t,t}} 
\sim (\tilde{\boldsymbol{M}}_{t}, \boldsymbol{B}_{t}) 
(\boldsymbol{H}_{t}, \tilde{\boldsymbol{Q}}_{t})^{\dagger}
\tilde{\boldsymbol{q}}_{t}
+ \boldsymbol{\Phi}_{(\tilde{\boldsymbol{M}}_{t}, \boldsymbol{B}_{t})}^{\perp}
\boldsymbol{\omega}_{t},
\end{equation}
with 
\begin{equation}
\boldsymbol{\omega}_{t} 
= \tilde{\boldsymbol{V}}
(\boldsymbol{\Phi}_{(\boldsymbol{H}_{t}, \tilde{\boldsymbol{Q}}_{t})}^{\perp})^{\mathrm{T}} 
\tilde{\boldsymbol{q}}_{t}, 
\end{equation}
where $\tilde{\boldsymbol{V}}$ is an independent and Haar-distributed 
orthogonal matrix. Evaluating the pseudo-inverse matrix 
$(\boldsymbol{H}_{t}, \tilde{\boldsymbol{Q}}_{t})^{\dagger}$, we have 
\begin{IEEEeqnarray}{rl}
\boldsymbol{V}^{\mathrm{T}}\tilde{\boldsymbol{q}}_{t}|_{\mathfrak{E}_{t,t}} 
\sim& \boldsymbol{B}_{t}\boldsymbol{\beta}_{t} 
+ \boldsymbol{\Phi}_{(\tilde{\boldsymbol{M}}_{t}, \boldsymbol{B}_{t})}^{\perp}
\boldsymbol{\omega}_{t} \nonumber \\
&+ \tilde{\boldsymbol{M}}_{t}\boldsymbol{o}(1) 
+ \boldsymbol{B}_{t}\boldsymbol{o}(1), 
\end{IEEEeqnarray}
with 
\begin{equation}
\boldsymbol{\beta}_{t} 
= (\tilde{\boldsymbol{Q}}_{t}^{\mathrm{T}}
\tilde{\boldsymbol{Q}}_{t})^{-1} 
\tilde{\boldsymbol{Q}}_{t}^{\mathrm{T}}\tilde{\boldsymbol{q}}_{t}.   
\end{equation}

In order to complete the proof of Property~\ref{property_a_A} for $\tau=t$, 
we analyze the moment properties of $\boldsymbol{\omega}_{t}$. 
By definition, we have 
\begin{equation}
\|\boldsymbol{\omega}_{t}\|^{2} 
= \tilde{\boldsymbol{q}}_{t}^{\mathrm{T}}
\boldsymbol{P}_{(\boldsymbol{H}_{t},\tilde{\boldsymbol{Q}}_{t})}^{\perp}
\tilde{\boldsymbol{q}}_{t}
\aeq  \tilde{\boldsymbol{q}}_{t}^{\mathrm{T}}
\boldsymbol{P}_{\tilde{\boldsymbol{Q}}_{t}}^{\perp}\tilde{\boldsymbol{q}}_{t} 
+ o(N), 
\end{equation}
where the second equality follows from the orthogonality between 
$\boldsymbol{h}_{\tau'}$ and $\tilde{\boldsymbol{q}}_{\tau''}$. 
Furthermore, it is straightforward to confirm that $\boldsymbol{\omega}_{t}$ 
has bounded $(4+\epsilon)$th moments. Thus, Property~\ref{property_a_A} 
holds for $\tau=t$. 

See \cite{Takeuchi171} for the proof of Properties~\ref{property_b_A} and 
\ref{property_d_A} for $\tau=t$. Finally, we prove Property~\ref{property_c_A} 
for $\tau=t$. 
Let $\{\boldsymbol{z}_{\tau}\sim\mathcal{N}(\boldsymbol{0}, 
v_{\tau}\boldsymbol{I}_{N})\}$ denote independent Gaussian random vectors 
with $v_{0}=\lim_{M=\delta N\to\infty}N^{-1}\|\boldsymbol{q}_{0}\|^{2}$ and 
$v_{\tau}=\lim_{M=\delta N\to\infty}
N^{-1}\|\tilde{\boldsymbol{q}}_{\tau}^{\perp}\|^{2}$ for $\tau>0$, and define 
$\tilde{\boldsymbol{b}}_{\tau}$ recursively as 
\begin{equation}
\tilde{\boldsymbol{b}}_{\tau} 
= \tilde{\boldsymbol{B}}_{\tau}\boldsymbol{\beta}_{\tau} 
+ \boldsymbol{z}_{\tau}, 
\quad 
\tilde{\boldsymbol{B}}_{\tau} 
=(\tilde{\boldsymbol{b}}_{0},\ldots,\tilde{\boldsymbol{b}}_{\tau-1}), 
\end{equation}
conditioned on $\tilde{\boldsymbol{Q}}_{t+1}$, 
with $\tilde{\boldsymbol{b}}_{0}=\boldsymbol{z}_{0}$. 
Using Property~\ref{property_a_A} and Lemma~\ref{lemma_SLLN} repeatedly 
yields  
\begin{equation}
\frac{1}{N}\boldsymbol{b}_{\tau'}^{\mathrm{T}}\tilde{\boldsymbol{\phi}}_{t}
\aeq \frac{1}{N}\mathbb{E}\left[
 \tilde{\boldsymbol{b}}_{\tau'}^{\mathrm{T}}\tilde{\boldsymbol{\phi}}_{t}
 (\tilde{\boldsymbol{B}}_{t+1}, \tilde{\boldsymbol{w}})
\right] + o(1). 
\end{equation}
Since $\{\tilde{\boldsymbol{b}}_{\tau}\}$ are jointly Gaussian conditioned 
on $\tilde{\boldsymbol{Q}}_{t+1}$, we use Lemma~\ref{lemma_Stein} to obtain 
\begin{IEEEeqnarray}{rl}
\frac{1}{N}
\boldsymbol{b}_{\tau'}^{\mathrm{T}}\tilde{\boldsymbol{\phi}}_{t}
\aeq& \frac{1}{N}\sum_{t'=0}^{t}\mathbb{E}\left[
 \tilde{\boldsymbol{b}}_{\tau'}^{\mathrm{T}}\tilde{\boldsymbol{b}}_{t'}
\right]\mathbb{E}\left[
 \left\langle
  \partial_{t'}\tilde{\boldsymbol{\phi}}_{t}
  (\tilde{\boldsymbol{B}}_{t+1}, \tilde{\boldsymbol{w}})
 \right\rangle 
\right]
+ o(1) \nonumber \\
\aeq& \frac{1}{N}\sum_{t'=0}^{t}\tilde{\boldsymbol{b}}_{\tau'}^{\mathrm{T}}
\tilde{\boldsymbol{b}}_{t'}\left\langle
  \partial_{t'}\tilde{\boldsymbol{\phi}}_{t}
  (\boldsymbol{B}_{t+1}, \tilde{\boldsymbol{w}})
 \right\rangle 
+ o(1),  
\end{IEEEeqnarray}
where the last equality follows from the repetition of the argument in 
(\ref{bm0}). 
Thus, Property~\ref{property_c_A} holds for $\tau=t$. 

\subsection{Properties in Module B by Induction} 
Suppose that all properties in Modules A an B hold for all 
$\tau\leq t$ and $\tau<t$, respectively.  
It is possible to prove all properties in Module~B for $\tau=t$, 
by repeating the proof in Appendix~\ref{Module_A_induction}. 
Thus, Theorem~\ref{theorem_SE} is correct for all $\tau$. 

\section{Proof of Lemma~\ref{lemma_SLLN}}
\label{appen_proof_lemma_SLLN}
Since $\boldsymbol{a}\in\mathbb{R}^{N-t}$ is an orthogonally invariant vector, 
we can represent $\boldsymbol{a}$ as $\boldsymbol{a}
\sim\gamma\boldsymbol{u}_{1}$ with 
$\gamma=\|\boldsymbol{a}\|/\|\boldsymbol{u}_{1}\|$ and  
some standard Gaussian vector $\boldsymbol{u}
=(\boldsymbol{u}_{0}^{\mathrm{T}},\boldsymbol{u}_{1}^{\mathrm{T}})^{\mathrm{T}}\sim
\mathcal{N}(\boldsymbol{0}, \boldsymbol{I}_{N})$. For 
an $N\times N$ orthogonal matrix $\boldsymbol{\Phi}
=(\boldsymbol{\Phi}^{\parallel}, \boldsymbol{\Phi}^{\perp})$, let  
$\boldsymbol{z}=\boldsymbol{\Phi}\boldsymbol{u}
\sim\mathcal{N}(\boldsymbol{0},\boldsymbol{I}_{N})$ and 
$\boldsymbol{\epsilon} = \boldsymbol{\Phi}^{\parallel}\boldsymbol{u}_{0}$. 
Then, we have 
\begin{equation} \label{a_tilde}
\tilde{\boldsymbol{a}}
\sim \gamma(\boldsymbol{\Phi}\boldsymbol{u} 
- \boldsymbol{\Phi}^{\parallel}\boldsymbol{u}_{0})
= \gamma( \boldsymbol{z} - \boldsymbol{\epsilon}). 
\end{equation}
Note that $\gamma^{2}\ato v$ holds.  

Let 
\begin{equation}
S_{N}=\sum_{n=1}^{N}\tilde{f}_{n}(\tilde{a}_{n}), 
\end{equation}
\begin{equation}
\tilde{S}_{N}=\sum_{n=1}^{N}\tilde{f}_{n}(\gamma z_{n}), 
\end{equation}
\begin{equation}
\bar{S}_{N}=\sum_{n=1}^{N}\tilde{f}_{n}(\sqrt{v_{N}}z_{n}), 
\end{equation}
with $\tilde{f}_{n}(x)=f_{n}(x+b_{n})$ and $v_{N}=\|\boldsymbol{a}\|^{2}/(N-t)$. 
Since $f_{n}$ is pseudo-Lipschitz of $k$, $\tilde{f}_{n}$ is so. 
We first evaluate the difference $|\mathbb{E}[S_{N}]-\mathbb{E}[\bar{S}_{N}]|$. 
Using the triangle inequality yields 
\begin{equation}
|\mathbb{E}[S_{N}]-\mathbb{E}[\bar{S}_{N}]|
\leq |\mathbb{E}[S_{N}]-\mathbb{E}[\tilde{S}_{N}]|
+ |\mathbb{E}[\tilde{S}_{N}]-\mathbb{E}[\bar{S}_{N}]|. 
\end{equation}

We upper-bound the first term.  
From the pseudo-Lipschitz property of $\tilde{f}_{n}$, 
there is some constant $L>0$ such that   
\begin{IEEEeqnarray}{rl}
\left|
 \mathbb{E}[S_{N}] - \mathbb{E}[\tilde{S}_{N}]
\right| 
\leq& L\sum_{n=1}^{N}\mathbb{E}\left[
 |\gamma\epsilon_{n}|
\right]
+ L\sum_{n=1}^{N}\mathbb{E}\left[
 |\gamma z_{n}|^{k-1}|\gamma\epsilon_{n}|
\right] \nonumber \\
&+ L\sum_{n=1}^{N}\mathbb{E}\left[
 |\tilde{a}_{n}|^{k-1}|\gamma\epsilon_{n}|
\right]. 
\label{ES}
\end{IEEEeqnarray}

For the first term on the upper bound~(\ref{ES}), we use 
the upper bound $\|\boldsymbol{\epsilon}\|_{1}\leq\sqrt{N}
\|\boldsymbol{\epsilon}\|$ to obtain 
\begin{IEEEeqnarray}{rl}
&\sum_{n=1}^{N}\mathbb{E}\left[
 |\gamma\epsilon_{n}|
\right]  
\leq \sqrt{N}\mathbb{E}\left[
 |\gamma|\|\boldsymbol{\epsilon}\|
 \right] 
\leq \sqrt{N}\left(
 \mathbb{E}[\gamma^{2}]
 \mathbb{E}[\|\boldsymbol{\epsilon}\|^{2}]
\right)^{1/2} \nonumber \\
=& \sqrt{N}\left(
 \mathbb{E}[\gamma^{2}]
 \mathbb{E}[\|\boldsymbol{u}_{0}\|^{2}]
\right)^{1/2}
=\sqrt{N}\{\sqrt{vt}+o(1)\}, 
\end{IEEEeqnarray}
where the second inequality follows from the Cauchy-Schwarz inequality. 

For the second term on the upper bound~(\ref{ES}), similarly we have 
\begin{IEEEeqnarray}{rl}
&\sum_{n=1}^{N}\mathbb{E}\left[
 |\gamma z_{n}|^{k-1}|\gamma\epsilon_{n}|
\right] \nonumber \\
\leq& \mathbb{E}\left[
 |\gamma|^{k}\left(
  \sum_{n=1}^{N}|z_{n}|^{2k-2}
 \right)^{1/2}\|\boldsymbol{\epsilon}\|
\right] \nonumber \\ 
\leq& \left(
 \mathbb{E}[|\gamma|^{pk}]
\right)^{1/p}
\left\{
 \mathbb{E}\left[
 \left(
  \sum_{n=1}^{N}|z_{n}|^{2k-2} 
 \right)^{q/2} 
 \|\boldsymbol{\epsilon}\|^{q} 
\right]
\right\}^{1/q} \nonumber \\ 
\leq& C\left\{
 \mathbb{E}\left[
  \left(
   \sum_{n=1}^{N}|z_{n}|^{2k-2} 
  \right)^{q} 
 \right]
 \mathbb{E}\left[
  \|\boldsymbol{\epsilon}\|^{2q} 
 \right]
\right\}^{(2q)^{-1}}, 
\end{IEEEeqnarray}
for some constants $C>0$ and $q>1$. In these bounds, the first inequality 
follows from the Cauchy-Schwarz inequality. The second inequality is due to 
H\"older's inequality for all integers $p>1$, $q>1$, and $p+q\leq1$. The 
last inequality follows from the Cauchy-Schwarz inequality and 
$\mathbb{E}[|\gamma|^{pk}]<\infty$ for $p>1$ sufficiently close to $1$. 
The expectation $\mathbb{E}[\|\boldsymbol{\epsilon}\|^{2q}]
=\mathbb{E}[\|\boldsymbol{u}_{0}\|^{2q}]$ is bounded for fixed $t$. 
Furthermore, we use the boundedness of all moments of $|z_{n}|$ to have 
\begin{equation}
\left\{
 \mathbb{E}\left[
  \left(
   \sum_{n=1}^{N}|z_{n}|^{2k-2} 
  \right)^{q} 
 \right]
\right\}^{(2q)^{-1}}
< \left(
 C^{2q}N^{q}
\right)^{(2q)^{-1}}=C\sqrt{N} 
\end{equation}
for some constant $C>0$. Combining these observations, we arrive at 
\begin{equation}
\sum_{n=1}^{N}\mathbb{E}\left[
 |\gamma z_{n}|^{k-1}|\gamma\epsilon_{n}|
\right] < C\sqrt{N}
\end{equation}
for some constant $C>0$. 

For the last term on the upper bound~(\ref{ES}), we repeat the same argument 
to obtain 
\begin{equation}
\sum_{n=1}^{N}\mathbb{E}\left[
 |\tilde{a}_{n}|^{k-1}|\gamma\epsilon_{n}|
\right]
< C\sqrt{N} 
\end{equation} 
for some constant $C>0$. Thus, we have proved 
\begin{equation} \label{S_mean}
\frac{1}{N}|\mathbb{E}[S_{N}] - \mathbb{E}[\tilde{S}_{N}]| 
= {\cal O}(N^{-1/2}). 
\end{equation}

We upper-bound the difference 
$|\mathbb{E}[\tilde{S}_{N}]-\mathbb{E}[\bar{S}_{N}]|$. Using the 
pseudo-Lipschitz property of $\tilde{f}_{n}$ yields 
\begin{IEEEeqnarray}{rl}
|\mathbb{E}[\tilde{S}_{N}]-\mathbb{E}[\bar{S}_{N}]|
&\leq L\sum_{n=1}^{N}\mathbb{E}[|\gamma-\sqrt{v_{N}}||z_{n}|] 
\nonumber \\
+& L\sum_{n=1}^{N}\mathbb{E}\left[
 |\gamma|^{k-1}|\gamma-\sqrt{v_{N}}||z_{n}|^{k} 
\right] \nonumber \\ 
+& L\sum_{n=1}^{N}\mathbb{E}\left[
 v_{N}^{(k-1)/2}|\gamma-\sqrt{v_{N}}||z_{n}|^{k} 
\right] \label{ES_tilde}
\end{IEEEeqnarray}
for some constant $L>0$. For the second term on the upper 
bound~(\ref{ES_tilde}), we use the definitions 
$\gamma=\|\boldsymbol{a}\|/\|\boldsymbol{u}_{1}\|$ and 
$v_{N}=\|\boldsymbol{a}\|^{2}/(N-t)$ to obtain  
\begin{IEEEeqnarray}{rl}
&\sum_{n=1}^{N}\mathbb{E}[|\gamma|^{k-1}|\gamma-\sqrt{v_{N}}||z_{n}|^{k}] 
\nonumber \\
=& \mathbb{E}\left[
 \frac{|\gamma|^{k-1}|N-t-\|\boldsymbol{u}_{1}\|^{2}|\|\boldsymbol{a}\|
 \|\boldsymbol{z}\|_{k}^{k}}
 {\sqrt{N-t}\|\boldsymbol{u}_{1}\|(\sqrt{N-t}+\|\boldsymbol{u}_{1}\|)}
\right] \nonumber \\
<& \mathbb{E}\left[
 |\gamma|^{k-1}|N-t-\|\boldsymbol{u}_{1}\|^{2}|
 \frac{\|\boldsymbol{a}\|\|\boldsymbol{z}\|_{k}^{k}}
 {(N-t)\|\boldsymbol{u}_{1}\|}
\right] \nonumber \\
\leq& C\left(
 \mathbb{E}\left[
  |N-t-\|\boldsymbol{u}_{1}\|^{2}|^{q}
  \left(
   \frac{\|\boldsymbol{a}\|\|\boldsymbol{z}\|_{k}^{k}}
   {(N-t)\|\boldsymbol{u}_{1}\|}
  \right)^{q}
 \right]
\right)^{1/q} \nonumber \\ 
\leq& C\left(
 \mathbb{E}\left[
  |N-t-\|\boldsymbol{u}_{1}\|^{2}|^{2q}
 \right]
 \mathbb{E}\left[
  \left(
   \frac{\|\boldsymbol{a}\|\|\boldsymbol{z}\|_{k}^{k}}
   {(N-t)\|\boldsymbol{u}_{1}\|}
  \right)^{2q}
 \right]
\right)^{\frac{1}{2q}}  
\end{IEEEeqnarray}
for some constants $C>0$ and $q>1$. In these bounds, 
the second inequality follows from H\"older's inequality. The last 
inequality is due to the Cauchy-Schwarz inequality. 
Since $\boldsymbol{u}_{1}\sim\mathcal{N}(\boldsymbol{0},\boldsymbol{I}_{N-t})$ 
holds, it is straightforward to confirm  
\begin{equation}
\mathbb{E}[(N-t-\|\boldsymbol{u}_{1}\|^{2})^{2q}] 
=\mathbb{E}\left[
 \left\{
  \sum_{n=t+1}^{N}(u_{n}^{2}-1)
 \right\}^{2q}
\right]= {\cal O}(N^{q}).  
\end{equation}
Furthermore, the latter expectation on the upper bound is bounded. 
Thus, we arrive at 
\begin{equation}
\sum_{n=1}^{N}\mathbb{E}[|\gamma|^{k-1}|\gamma-\sqrt{v_{N}}||z_{n}|^{k}]
= {\cal O}(\sqrt{N}). 
\end{equation}

Repeating the same argument for the first and last terms on the upper 
bound~(\ref{ES_tilde}), we have 
\begin{equation} \label{S_mean_tilde}
|\mathbb{E}[\tilde{S}_{N}]-\mathbb{E}[\bar{S}_{N}]|
= {\cal O}(\sqrt{N}). 
\end{equation} 
Combining (\ref{S_mean}) and (\ref{S_mean_tilde}), we arrive at 
\begin{equation}
\frac{1}{N} |\mathbb{E}[\tilde{S}_{N}]-\mathbb{E}[\bar{S}_{N}]|
= {\cal O}(N^{-1/2}). 
\end{equation}

We next prove that $(S_{N}-\mathbb{E}[S_{N}])/N$ converges a.s.\  
to zero as $N\to\infty$. From a strong law of large numbers 
(SLLN) for dependent random variables~\cite[Corollary 1]{Lyons88}, 
it is sufficient to prove $\mathbb{V}[S_{N}]={\cal O}(N^{a})$ for some $a<2$. 

Let us prove $\mathbb{V}[S_{N}]=\mathbb{V}[\tilde{S}_{N}]+{\cal O}(N^{a})$ 
for $a<2$. By definition, we have 
\begin{equation}
S_{N}^{2} 
= \sum_{n,n'=1}^{N}\tilde{f}_{n}(\tilde{a}_{n})\tilde{f}_{n'}(\tilde{a}_{n'}). 
\end{equation}
Proposition~\ref{proposition3} implies that $S_{N}^{2}$ is the sum of 
the pseudo-Lipschitz functions $f(\tilde{a}_{n},\tilde{a}_{n'})
=\tilde{f}_{n}(\tilde{a}_{n})\tilde{f}_{n'}(\tilde{a}_{n'})$ of order $2k$. 
Thus, we have 
\begin{IEEEeqnarray}{rl}
\left|
 \mathbb{E}[S_{N}^{2}] - \mathbb{E}[\tilde{S}_{N}^{2}]
\right|
&< L\sum_{n,n'}\mathbb{E}\left[
 \gamma(\epsilon_{n}^{2} + \epsilon_{n'}^{2})^{1/2}
\right] \nonumber \\ 
+ L\sum_{n,n'}&\mathbb{E}\left[
 \gamma^{2k}(z_{n}^{2}+z_{n'}^{2})^{k-1/2}¡¡
 (\epsilon_{n}^{2} + \epsilon_{n'}^{2})^{1/2}
\right] \nonumber \\
+ L\sum_{n,n'}&\mathbb{E}\left[
 \gamma(\tilde{a}_{n}^{2}+\tilde{a}_{n'}^{2})^{k-1/2}
 (\epsilon_{n}^{2} + \epsilon_{n'}^{2})^{1/2} \label{ES2}
\right]\end{IEEEeqnarray}
for some constant $L>0$. 

For the first term on the upper bound~(\ref{ES2}), 
we use H\"older's inequality to obtain 
\begin{IEEEeqnarray}{l}
\sum_{n,n'}\mathbb{E}\left[
 \gamma(\epsilon_{n}^{2} + \epsilon_{n'}^{2})^{1/2}
\right]
\leq C\sum_{n,n'}\left\{
 \mathbb{E}\left[
  (\epsilon_{n}^{2} + \epsilon_{n'}^{2})^{q/2}
 \right]
\right\}^{1/q} \nonumber \\
\leq CN^{2}\left\{
 \mathbb{E}\left[
  \frac{1}{N^{2}}\sum_{n,n'}(\epsilon_{n}^{2} + \epsilon_{n'}^{2})^{q/2}
 \right]
\right\}^{1/q}
\end{IEEEeqnarray}
for some constants $C>0$ and $q\geq 2$, 
where the second inequality follows from Jensen's inequality. 
Applying the inequality $|x_{1}|+|x_{2}|\leq 2^{1-2/q}
(|x_{1}|^{q/2}+|x_{2}|^{q/2})^{2/q}$, we have 
\begin{IEEEeqnarray}{rl}
\sum_{n,n'}\mathbb{E}\left[
 \gamma(\epsilon_{n}^{2} + \epsilon_{n'}^{2})^{1/2}
\right]
\leq CN^{2}\left\{
 \mathbb{E}\left[
  \frac{1}{N}\sum_{n=1}^{N}|\epsilon_{n}|^{q}
 \right]
\right\}^{1/q}
\end{IEEEeqnarray}
for some constant $C>0$. Since 
$\sum_{n=1}^{N}|\epsilon_{n}|^{q}\leq\|\boldsymbol{\epsilon}\|^{q}
=\|\boldsymbol{u}_{0}\|^{q}$ holds, we arrive at 
\begin{equation}
\sum_{n,n'}\mathbb{E}\left[
 \gamma(\epsilon_{n}^{2} + \epsilon_{n'}^{2})^{1/2}
\right]
\leq CN^{2-1/q}
\end{equation}
for some constant $C>0$. 

For the second term on the upper bound~(\ref{ES2}), similarly we have 
\begin{IEEEeqnarray}{rl}
&\sum_{n,n'}\mathbb{E}\left[
 \gamma^{2k}(z_{n}^{2}+z_{n'}^{2})^{k-1/2}¡¡
 (\epsilon_{n}^{2} + \epsilon_{n'}^{2})^{1/2}
\right] \nonumber \\
\leq& CN^{2}\left\{
 \mathbb{E}\left[
  \frac{1}{N}\sum_{n=1}^{N}\epsilon_{n}^{q}
 \right]
\right\}^{1/q} ={\cal O}(N^{2-1/q})
\end{IEEEeqnarray}
for some constants $C>0$ and $q\geq2$, where we have used 
$\mathbb{E}[\gamma^{2pk}]<\infty$ and 
$\mathbb{E}[(z_{n}^{2}+z_{n'}^{2})^{q(2k-1)}]<\infty$ for some $q$ and 
$p=(1-q^{-1})^{-1}$. Repeating the same argument 
for the last term on the upper bound~(\ref{ES2}), we arrive at 
\begin{equation}
\left|
 \mathbb{E}[S_{N}^{2}] - \mathbb{E}[\tilde{S}_{N}^{2}]
\right|
= {\cal O}(N^{2-1/q})
\end{equation}
for some $q\geq2$. 
Since (\ref{S_mean}) implies $(\mathbb{E}[S_{N}])^{2}
=(\mathbb{E}[\tilde{S}_{N}])^{2}+{\cal O}(N^{3/2})$, we have 
$\mathbb{V}[S_{N}]=\mathbb{V}[\tilde{S}_{N}]+{\cal O}(N^{2-1/q})$ for some 
$q\geq2$. 

In order to prove the SLLN for $(S_{N}-\mathbb{E}[\bar{S}_{N}])/N$, we need to 
show $|\mathbb{E}[\tilde{S}_{N}^{2}]-\mathbb{E}[\bar{S}_{N}^{2}]|
={\cal O}(N^{a})$ for some $a<2$. This convergence and (\ref{S_mean_tilde}) 
imply that $N^{-1}(\mathbb{E}[\tilde{S}_{N}]-\mathbb{E}[\bar{S}_{N}])\to0$ and 
$\mathbb{V}[\tilde{S}_{N}]=\mathbb{V}[\bar{S}_{N}]+{\cal O}(N^{\max\{3/2,a\}})$.   
Furthermore, it is straightforward to confirm 
\begin{equation}
\mathbb{V}\left[
 \bar{S}_{N}
\right] 
= \sum_{n=1}^{N}\mathbb{E}\left\{
 \mathbb{V}\left[
  \left.
   \tilde{f}_{n}(\sqrt{v}_{N} z_{n})
  \right| \|\boldsymbol{a}\|
 \right]
\right\}
= {\cal O}(N).  
\end{equation}
Thus, we find the SLLN $(S_{N}-\mathbb{E}[\bar{S}_{N}])/N\ato0$.  

Let us prove $|\mathbb{E}[\tilde{S}_{N}^{2}]-\mathbb{E}[\bar{S}_{N}^{2}]|
={\cal O}(N^{a})$ for some $a<2$. Using the pseudo-Lipschitz property yields 
\begin{IEEEeqnarray}{rl}
&|\mathbb{E}[\tilde{S}_{N}^{2}]-\mathbb{E}[\bar{S}_{N}^{2}]| 
\nonumber \\ 
\leq& L\sum_{n,n'}\mathbb{E}\left[
 |\gamma-\sqrt{v_{N}}|(z_{n}^{2}+z_{n'}^{2})^{1/2}
\right] \nonumber \\
&+ L \sum_{n,n'}\mathbb{E}\left[
 \gamma^{2k-1}(z_{n}^{2} + z_{n'}^{2})^{k}
 |\gamma-\sqrt{v_{N}}|
\right] \nonumber \\
&+ L \sum_{n,n'}\mathbb{E}\left[
 v_{N}^{k-1/2}(z_{n}^{2} + z_{n'}^{2})^{k}
 |\gamma-\sqrt{v_{N}}|
\right] \label{ES2_tilde}
\end{IEEEeqnarray}
for some constant $L>0$. For the second term on the upper 
bound~(\ref{ES2_tilde}), we use the definitions $\gamma=\|\boldsymbol{a}\|
/\|\boldsymbol{u}_{1}\|$ and $v_{N}=\|\boldsymbol{a}\|^{2}/(N-t)$ to obtain 
\begin{IEEEeqnarray}{rl}
&\sum_{n,n'}\mathbb{E}\left[
 \gamma^{2k-1}(z_{n}^{2} + z_{n'}^{2})^{k}
 |\gamma-\sqrt{v_{N}}|
\right] \nonumber \\
<& \sum_{n,n'}\mathbb{E}\left[
 \gamma^{2k-1}(z_{n}^{2} + z_{n'}^{2})^{k}
 \frac{|N-t-\|\boldsymbol{u}_{1}\|^{2}|\|\boldsymbol{a}\|}
 {(N-t)\|\boldsymbol{u}_{1}\|}
\right] \nonumber \\
\leq& CN\left\{
 \mathbb{E}\left[
  \left(
   |N-t-\|\boldsymbol{u}_{1}\|^{2}|
   \frac{\|\boldsymbol{a}\|\|\boldsymbol{z}\|_{2k}^{2k}}
   {(N-t)\|\boldsymbol{u}_{1}\|}
  \right)^{q}
 \right]
\right\}^{1/q} \nonumber \\
=& {\cal O}(N^{3/2})
\end{IEEEeqnarray}
for some constants $C>0$ and $q>1$, where the second inequality follows from 
$z_{n}^{2}+z_{n'}^{2}\leq 2^{1-1/k}(z_{n}^{2k}+z_{n'}^{2k})^{1/k}$, 
H\"older's inequality, and from 
$\mathbb{E}[\gamma^{p(2k-1)}]<\infty$ for $p=(1-q^{-1})^{-1}$. 
Repeating the same argument for the first and last terms on the upper 
bound~(\ref{ES2_tilde}), we arrive at 
\begin{equation}
|\mathbb{E}[\tilde{S}_{N}^{2}]-\mathbb{E}[\bar{S}_{N}^{2}]|
={\cal O}(N^{3/2}). 
\end{equation}

Let $S_{0,N}=\sum_{n=1}^{N}f_{n}(\sqrt{v}z_{n})$.  
In order to complete the proof of Lemma~\ref{lemma_SLLN}, 
we show $N^{-1}|\mathbb{E}[\bar{S}_{N}] - \mathbb{E}[S_{0,N}]|\to0$. 
Define $\bar{S}_{0,N}=\sum_{n=1}^{N}\tilde{f}_{n}(\sqrt{v}z_{n})$. 
Using the triangle inequality yields 
\begin{IEEEeqnarray}{rl}
&|\mathbb{E}[\bar{S}_{N}] - \mathbb{E}[S_{0,N}]| 
\nonumber \\
\leq& |\mathbb{E}[\bar{S}_{N}] - \mathbb{E}[\bar{S}_{0,N}]|  
+ |\mathbb{E}[\bar{S}_{0,N}] - \mathbb{E}[S_{0,N}]|. 
\end{IEEEeqnarray}
It is straightforward to prove $N^{-1}|\mathbb{E}[\bar{S}_{N}] 
- \mathbb{E}[\bar{S}_{0,N}]|\to0$. Thus, we only evaluate the second term. 
 
Using the pseudo-Lipschitz property yields 
\begin{IEEEeqnarray}{rl}
&\frac{1}{N}\left|
 \mathbb{E}[\bar{S}_{0,N}]
 - \mathbb{E}[S_{0,N}]  
\right| \nonumber \\ 
\leq& \frac{L}{N}\sum_{n=1}^{N}|b_{n}|\mathbb{E}_{z_{n}}\left[
  1 + (b_{n}+\sqrt{v}z_{n})^{k-1} + (\sqrt{v}z_{n})^{k-1}
\right] \nonumber \\
\leq& L\left(
 \frac{C_{N}}{N}\|\boldsymbol{b}\|^{2}
\right)^{1/2} 
\end{IEEEeqnarray}
for some constant $L>0$, with 
\begin{equation}
C_{N}= \frac{1}{N}\sum_{n=1}^{N}\left(
 \mathbb{E}_{z_{n}}\left[
  1 + (b_{n}+\sqrt{v}z_{n})^{k-1} + (\sqrt{v}z_{n})^{k-1}
 \right]
\right)^{2}, 
\end{equation}
where the second inequality follows from the Cauchy-Schwarz inequality,  
Since $N^{-1}\|\boldsymbol{b}\|^{2}\to0$ and $N^{-1}\sum_{n=1}^{N}b_{n}^{2k-2}
<\infty$ are assumed, we arrive at 
$N^{-1}|\mathbb{E}[\bar{S}_{0,N}] - \mathbb{E}[S_{0,N}]|\ato0$. 
Thus, Lemma~\ref{lemma_SLLN} holds.

\end{document}